\documentclass[%
 reprint,
superscriptaddress,
nofootinbib,
 amsmath,amssymb,
 aps,
 prd,
]{revtex4-2}

\usepackage{graphicx}
\usepackage{dcolumn}
\usepackage{bm}
\usepackage{comment}

\usepackage[normalem]{ulem}
\usepackage{xcolor}
\usepackage{cancel}

\newcommand{\Eq}[1]{Eq.~(\ref{#1})}
\newcommand{\Sec}[1]{Section~\ref{#1}}

\begin{document}


\title{Bondi-Hoyle-Lyttleton accretion onto ultra dense dark matter halos and
direct collapse black holes}

\author{Kandaswamy Subramanian}%
\email{kandu@iucaa.in;kandaswamy.subramanian@ashoka.edu.in}
\affiliation{Department of Physics, Ashoka University, Rajiv Gandhi Education City, Rai, Sonipat 131029, Haryana, India}
\affiliation{%
IUCAA, Post Bag 4, Ganeshkhind, Pune 411007, India}

\author{Bikram Phookun}
\email{bikram.phookun@ashoka.edu.in}
\affiliation{Department of Physics, Ashoka University, Rajiv Gandhi Education City, Rai, Sonipat 131029, Haryana, India}





\date{\today}

\begin{abstract}

We suggest a formation scenario of black holes with
intermediate mass $\sim 10^3 M_\odot$, by post recombination Bondi-Hoyle-Lyttleton 
accretion into ultra dense dark matter halos (UDMH)
of $\sim 10^5 M_\odot$, which have formed around the recombination epoch. Such UDMH can result
from rare curvature fluctuations on small scales whose amplitude is still well below the current Cosmic Microwave Background 
(CMB) spectral distortion
limits. Gas accreted by the UDMH is heated to virial temperatures above which
atomic cooling is efficient, cools rapidly to about $\sim 8000$ K and collapses on the free fall time of few $10^4$ yr
to the halo core, until supported by rotation.
Further fragmentation due to molecular cooling is prevented by the suppression of $H_2$ molecule
formation by the CMB photons at redshifts $z> 200-400$. We find that
the rotationally supported gas disk will be compact and massive enough 
to undergo self-gravitational instability
in some cases, plausibly where accretion is into a nearly spherical UDMH which 
has formed from a rare peak in the density field.
This results in a further, rapid transfer of mass inwards due to viscous forces and gravitational torques 
leading to the formation of a supermassive star and/or
black hole of about $10^3 M_\odot$ at redshifts of a few hundred. 
Such intermediate mass black holes formed at high redshifts can have a large-enough abundance to
seed the first super massive black holes and help explain the abundance of active galaxies
detected now at increasingly larger redshifts by the James Webb Space Telescope.

\end{abstract}

\maketitle


\section{\label{sec:intro}Introduction}

In the standard model of cosmology, galaxies and other large-scale structures
form by the growth and collapse of initially small density fluctuations
amplified by self-gravitational instability. Such a picture is supported by a wide
spectrum of observations. These include the precise determination of the curvature power spectrum 
from the observed Cosmic Microwave Background (CMB) anisotropies \cite{Planck18} 
and the Lyman-$\alpha$ forest data \cite{Bird+,Fernandez+}. These observations however  only determine 
the power spectrum of fluctuations on comoving length scales larger than about a Mpc.
Power on smaller length scales is much more weakly constrained by direct observations 
(see \cite{Bringmann+} for a recent summary and references). In principle, it can be 
much larger than that naive extrapolation from the large-scale
data. Such enhanced power will lead to earlier formation of collapsed small-scale structures and
perhaps novel phenomena, which can in turn constrain the small-scale
power spectrum or even explain current puzzles. 

Indeed, there are several intriguing features related to observations of 
black holes in the universe. The frequency of LIGO-VIRGO-KAGRA (LVK) detection of
gravitational waves from merging black holes in the mass range $10-100 M_\odot$ led to a surge of 
interest whether some of these black holes could be of primordial origin \cite{Carr_Green}. 
In some cases like the recent event GW231123, the inferred 
progenitor black hole masses are of order $100 M_\odot$, and lie in a mass gap $60-130 M_\odot$,
which is difficult to explain as a stellar remnant 
\cite{LVKGW231123}.   
This could indicate the role of primordial black holes (PBH), although hierarchical
mergers of stellar mass black holes is another possibility.

Moreover, the James Webb Space Telescope (JWST) has detected an abundance
of high redshift Active Galactic nuclei (AGNs) where the black hole mass to stellar
mass is higher than what obtains in the local universe \cite{Harikane+,Ubler+, Maiolino+,Volonteri25}.
This seems to suggest an early black hole formation before the bulk of the stellar mass, instead
of co-evolving together with the galaxy \cite{Delos_Silk}.
Some black hole candidates, like UHZ1 \cite{Priya24,Bogdan+} and GHZ9 \cite{Kovacs+} have such high masses $\sim 10^8 M_\odot$
by high redshifts $z \sim 10$, that black hole seeds need to be much 
heavier than expected from
stellar evolution. For this they need to be perhaps primordial
or at least form at sufficiently high redshifts \cite{Paccuci15,Paccuci23,Trinca2023,Dayal24,Jeon25a,Jeon25b,Volonteri25}.
An alternative possibility for explaining these high-redshift AGNs detected by
the JWST is of course, super-Eddington accretion, which does not necessarily need heavy seeds 
\cite{Paccuci15,fei2026}.

We explore here the possibility that very high redshift massive black holes seeds 
arise from baryonic accretion onto ultradense dark matter halos (UDMH). 
Note that UDMH may form in abundance even before recombination, from more typical density
fluctuations in models where rarer large-amplitude density fluctuations collapse 
to form primordial black holes \cite{Delos_Silk,FF25}. The formation of UDMH may also
arise in inflationary models, where curvature fluctuations are not strong enough to lead to
primordial black holes, but nevertheless are large enough that UDMH 
collapse by recombination. In the next section we
consider the formation of UDMH, constrained by the upper limit to the small
scale power spectrum set by Cosmic Microwave Background (CMB) observations \cite{COBE_FIRAS}. Before recombination, 
baryonic accretion onto a UDMH is 
strongly suppressed by the radiative viscosity and baryon streaming velocity relative to dark matter.
This is quantified by a linear theory calculation in Appendix~\ref{sec:baryon1}.
\Sec{sec:baryon2} 
then examines the Bondi-Hoyle-Littleton accretion post recombination.
The possibility of direct collapse black hole formation is considered in 
\Sec{sec:DCBH} 
and
our conclusions are presented in the last section.

\section{\label{sec:UDMH}Ultradense dark matter halos}

We discuss briefly the formation of ultradense dark matter halos (UDMH)
to set the scene for the subsequent sections where we consider the baryonic response. 
We mainly follow \cite{Delos_Silk} who have treated this issue in some detail. 
The scales (or wavenumber $k$) corresponding to primordial 
curvature perturbation $\zeta(k)$, 
\footnote{
The $k$ dependent quantities in this subsection, for example $\zeta(k)$ 
represent the variance of the corresponding (curvature) perturbation smoothed on
the mass scale $M = 6\pi^2 \rho_{d0} k^{-3}$, 
which is evaluated using a sharp-$k$ space filter \cite{Benson+12}.
Here $\rho_{d0}$ is the present day average dark matter density.}
which 
collapses to form UDMH are expected to enter the
horizon deep in the radiation dominated era when the scale factor 
is \cite{Hu_Sugiyama}, $a_H = 2^{-1/2} (k_{eq}/k) a_{eq}$. Here $a_{eq} \approx 3 \times 10^{-4}$
and $k_{eq} \approx 0.01$ Mpc$^{-1}$ are respectively the scale factor and Horizon scale at matter radiation equality.
We adopt a matter density parameter $\Omega_{m} h^2 = 0.14$ \cite{Planck18},
with $h=0.7$ the Hubble constant in units of $100$ km s$^{-1}$ Mpc$^{-1}$.
The subsequent growth of density
perturbations in the radiation-dominated era is given by  
$\delta(a,k) = I_1 \zeta(k) \ln(I_2 (a/a_H))$, with
$I_1\approx 6.4$ and $I_2\approx 0.47$ \cite{Hu_Sugiyama,Delos_Silk}.
A semi-analytic match to the matter dominated era for small scales ($k\gg k_{eq}$) is also
given in \cite{Hu_Sugiyama} (Eq. D3), 
\begin{equation}
   \delta(a,k) = I_1 \ \zeta(k) \ \ln \left(4 \ I_2 e^{-3} \frac{a_{eq}}{a_H}\right)\left(1+ \frac{3}{2} \frac{a}{a_{eq}}\right). 
   \label{HuSug}
\end{equation}
This is valid for small baryon fraction and also neglects the very small 
contribution from the decaying mode. We can use \Eq{HuSug} to estimate whether a
given mass of UDMH can collapse by recombination.

A perturbation of comoving wavenumber $k$ is associated with an UDMH of mass 
$M = 6\pi^2 \rho_{d0} k^{-3} \approx 2 \times 10^{12} M_\odot (k/{\rm Mpc}^{-1})^{-3}$ \cite{Lacey_Cole},
where we have also adopted a dark matter density parameter $\Omega_{d} h^2 = 0.12$ \cite{Planck18},
with $h=0.7$.
This also gives a present day dark matter density $\rho_{d0} =
3.3 \times 10^{10} M_\odot/{\rm Mpc}^3$. Then a UDMH with $M=10^5 M_\odot$
is associated with a comoving wavenumber $k \approx 270$ Mpc$^{-1}$, enters the Horizon
at a scale factor $a_H \approx 7.8 \times 10^{-9}$.
Using \Eq{HuSug} such a perturbation grows
by a factor of $\sim 130$ by $a = a_{eq}$,
a factor of $\sim 290$ by the epoch of recombination 
and a factor $\sim 470$ by a post recombination redshift of $z=600$. Thus an initial 
curvature perturbation 
of $\zeta \sim 5.8 \times 10^{-3}$ would have grown to $\delta =\delta_c \sim 1.686$,
become nonlinear and collapsed by the epoch of recombination.
For a Gaussian distribution of curvature perturbations, if we want only rarer $n\sigma$ perturbation to collapse
by recombination, we require smaller level of $\zeta =5.8/n \times 10^{-3}$.
For collapse by $z=600$ (a redshift we also consider below), one requires an even
smaller $\zeta = 3.56/n \times 10^{-3}$.
Here for illustration we have taken the same threshold $\delta_c$ for collapse as
in the spherical model, as would be relevant for collapse near recombination where the universe becomes
matter dominated. For a $10^6 M_\odot$ UDMH, the corresponding growth
factor from its entering the Hubble radius to recombination (using \Eq{HuSug}) is $\sim 260$. Thus these more massive
halos require a slightly larger $\zeta \sim 6.5 \times 10^{-3}$ to collapse by recombination.

Delos and Silk \cite{Delos_Silk} consider perturbations with much
larger $\zeta$, following a model by \cite{Carr+}, in which a significant fraction 
in UDMH up to $10^5 M_\odot$ collapse by $a_{eq}$.
However too large a power on small scales, will lead to detectable spectral distortions of the
CMB, due to energy input from Silk
damping of small-scale baryon photon acoustic oscillations.
The current upper limits by COBE on CMB spectral distortions, 
give an upper limit of $9\times 10^{-5}$ on the $\mu$-type distortions
\cite{COBE_FIRAS}, which
for delta-function perturbation spectra, translate to an upper limit
of $\zeta \approx 6.4 \times 10^{-3}$ (using Eq. 8 of \cite{Nakama_Carr_Silk})
on these small scales. Thus the $10^5 M_\odot$ UDMH considered by \cite{Delos_Silk}
is severely constrained.
For Gaussian curvature spectra which satisfy the COBE constraint, 
even typical fluctuations of $10^5-10^6 M_\odot$, can collapse to form UDMH 
by the epoch of recombination.
If we require only the rare UDMH of $10^5 M_\odot$, say corresponding to a $5\sigma-6\sigma$ 
perturbation, to collapse by recombination, 
the RMS power in curvature perturbations required as noted above is
$\zeta \sim 0.97-1.16 \times 10^{-3}$, comfortably below the WMAP limit.
\footnote{
For collapse of these rare $5\sigma-6\sigma$ perturbations by $z\sim 600$, one requires an even smaller 
$\zeta \sim 6-7 \times 10^{-4}$, which would be consistent with \cite{GR24}, where constraints $\sim 10$ times 
smaller in $\zeta$ are claimed, from the sizes of a few ultra-faint dwarf galaxies.}
Curvature perturbations for both more massive UDMH and those smaller than about $10^3 M_\odot$ are more tightly 
constrained (cf \cite{Bringmann+} and references therein), although rarer fluctuations on these mass scales could still collapse by recombination.

The comoving number density of UDMH above a mass $M$ is given using Press-Schechter theory \cite{PS} by
$N(>M,z) = (\bar{\rho}/M) {\rm erfc}[\delta_c/\sqrt{2}\sigma(M,z)]$, where ${\rm erfc}(x)$ is the complementary error function, 
and $\sigma(M,z)$ is the variance of the fractional density contrast at redshift $z$.
If the extra power on small scales is peaked around $10^5 M_\odot$, then the
comoving abundance of even rare $5\sigma-6\sigma$ collapsed halos is given by
$N \sim 1.6 \times 10^{-1}$ Mpc$^{-3}$ to 
$N \sim 6.3 \times 10^{-4}$ Mpc$^{-3}$
comparable to galactic abundances.

We can estimate several useful properties of the
virialized UDMH, using the spherical top-hat model and assuming collapse in the
matter dominated era at a redshift $z_{vir}$. Adopting $h=0.7$, these halos have \cite{BL01} a virial radius,
\begin{equation}
r_{vir} = 0.9 \left[ \frac{M}{10^5 \, M_\odot} \right]^{1/3} \left[ \frac{1+z_{vir}}{1100}\right]^{-1} \, \mathrm{pc}, 
\label{rvir}
\end{equation}
a circular velocity,
\begin{equation}
    v_c = 21.8 \left[ \frac{M}{10^5 \, M_\odot} \right]^{1/3} \left[ \frac{1+z_{vir}}{1100}\right]^{1/2} \, \mathrm{km \ s^{-1}}, 
    \label{vc}
\end{equation}
and virial temperature
\begin{equation}
T_{vir} = 1.7 \times 10^4 \, \left[ \frac{M}{10^5 \, M_\odot} \right]^{2/3} \, \left[ \frac{1+z_{vir}}{1100} \right] \, K.
\label{Tvir}
\end{equation}

Thus they are halos where atomic cooling is possible, which plays a crucial role as we will see below.
We consider in Appendix~\ref{sec:baryon1},
the response of the baryons to the growing dark matter perturbations before recombination
and show that it is strongly suppressed by radiation drag.
After recombination the baryons become predominantly neutral and friction due to Compton
scattering by the CMB photons becomes inefficient. Significant accretion becomes possible
to which we now turn. 

\section{\label{sec:baryon2}Bondi-Hoyle-Lyttleton Process}

We saw in \Sec{sec:UDMH}
that a virialized ultra-dense dark-matter halo (UDMH)  with mass $10^5 M_\odot$ can form by recombination. In this section we look at the accretion of baryons onto such a UDMH. The central question we seek to answer is the following: is it possible for the baryons to accrete onto the UDMH in such a way as to form a direct-collapse black hole? 

The virial radius of such a UDMH formed at $z = 1100$ \cite{BL01} is of the order of a parsec. 
One might imagine the surrounding baryonic matter to be stationary with respect to the halo, but,
as shown in \cite{Tsel_2010}, the baryonic matter at this time is typically streaming with respect to the 
dark matter. In the rest frame of the dark matter halo this steaming velocity has a Gaussian probability 
distribution, with a variance of about $30$ km s$^{-1}$, whereas the speed of sound in the baryons after recombination is about $6$ km s$^{-1}$, i.e. the baryonic matter is streaming at supersonic velocities past the UDMH. In these circumstances, the way in which the baryons accrete onto the UDMH may be described by the Bondi-Hoyle-Lyttleton (BHL) process.

Bondi \cite{Bondi_1952} looked at the problem of spherically-symmetric, steady-state accretion of gas onto a point mass. Hoyle and Lyttleton \cite{Hoyle_Lyt_1939} on the 
other hand considered the case of a point object moving through the ambient gas. The major difference is that the fluid elements now effectively follow hyperbolic streamlines around the point object when seen from its frame. The orbits of the gas flowing in from different directions collide downstream of the object along the axis of symmetry, which is along the direction of asymptotic
relative velocity between the accreting point mass and the gas. In the Hoyle-Lyttleton picture, the gas out to a certain maximum impact parameter -- called the accretion radius 
$r_{acc} = 2 GM/v_{\infty}^2$,
where $M$ is the mass of the object and $v_{\infty}$ the velocity of the ambient gas with respect to it at large distances -- is effectively funneled post-collision towards the central object. The process and its generalization to also include the Bondi accretion, known as Bondi-Hoyle-Lyttleton (BHL) accretion, is reviewed in 
\cite{Edgar_2004}, and a diagram showing the process can be found in \cite{OHSUGI201844}. 

The accretion in a realistic case is of course not onto a point object. For example, in our case, the UDMH is characterized by its virial radius. However, for a fixed UDMH mass, the virial radius remains constant after the halo forms, whereas the accretion radius, which depends on the streaming velocity of the gas with respect to the central object, grows as as $(1+z)^{-2}$, since the streaming velocity falls as $v_\infty(z) \propto 1+z$ \cite{Tsel_2010}. 

Although both virial radius and the accretion radius are similar at the epoch of virialization, the ratio of the latter to the former grows as $(1+z)^{-2}$, and the accretion radius rapidly becomes much larger that the virial radius, making the virialized UDMH effectively point-like from the point of view of accretion.

Let us also consider what happens if the halo grows during the process of accretion. As shown in \cite{Bert}, in this case the mass of the UDMH will grow as $(1 + z)^{-1}$. The process of virialization is now continuous and $z_{vir}$ is no longer fixed. From \Eq{rvir}, we see that $r_{vir} \propto (1+z)^{-4/3}$, which agrees with the growth rate for the halo quoted in \cite{Bert}. On the other hand, since $r_{acc}$ depends on $M (1+z)^{-2}$ (as shown in \Eq{racc} below) and therefore grows as $(1+z)^{-7/3}$. The ratio of $r_{acc}$ to $r_{vir}$ now grows as $(1+z)^{-1}$, and the point approximation for the halo remains reasonable.

A final point to note is that the accretion rate grows as $(1+z)^{-5/2}$ for the case of the fixed UDMH mass, and as $(1+z)^{-9/2}$ when the mass of the UDMH grows as $(1+z)^{-1}$ (see \Eq{dmdz_mfixed} and \Eq{dmdz_mvar} below). In other words, most of the accretion takes place at the later stages of the period of accretion, by which time $r_{acc}$ is much larger than $r_{vir}$.

We will therefore assume the point-mass accretion model for simplicity.

\subsection{Gas Flow and Shocks}

The meeting of opposing gas trajectories downstream of the central object will give rise to shocks, leading to changes in density and temperature. The final post-shock density and temperature will depend in addition on the cooling of the shocked gas. In the case of the systems we are looking at, the immediate post-shock temperature (before cooling kicks in) can be calculated from the shock condition 
\begin{align}
    T =& \frac{3}{16} \frac{\mu m_p}{k_B} v_\infty^2(z) 
    = \frac{3}{16} \frac{\mu m_p}{k_B} v_{\infty}^2(z_{R}) \left( \frac{1+z}{1100} \right)^2 \nonumber \\
    =& 2.0 \times 10^4 \, \mu \left( \frac{v_{\infty}(z_R)}{ 30 \, \mathrm{km \, s^{-1}}}\right)^2 \left( \frac{1+z}{1100} \right)^2 \, \mathrm{K},
    \label{post-shock_temp}
\end{align}
where $\mu=0.6$ and $\mu=1.23$ for ionized and neutral gas of primordial composition \cite{Mo_White}.

With decreasing redshift, therefore, the post-shock temperature falls, and by about $z=800$ it drops to $10^4$ K and to less than $3000$ K at $z = 400$. The rate of atomic cooling, which dominates, has a peak at about $2.0 \times 10^4$ K, and drops by several orders of magnitude by $8000$ K (see \cite{BL01}). As we shall see below, when we look at cooling in somewhat greater detail, this implies that gas that is heated above $8000$ K cools down to about $8000$ K, whereas gas that is heated to below that temperature remains at its original temperature. 

\subsection{Ionization Fraction}

The ionization fraction of the gas flowing into the UDMH is essentially that of the primordial gas, and will vary with redshift, being close to neutral immediately after recombination and with an ionization fraction of about $10^{-4}$ at $z = 400$. After the shock, the ionization fraction will in general change, and will be determined by the balance between collisional ionization and radiative recombination which leads to (\cite{Draine_book}):
\begin{equation}
   \frac{n_{HII}}{n_{HI}} = \frac{\alpha_{ci}}{\alpha_{rr}},
\end{equation}
where $\alpha_{ci}$ and $\alpha_{rr}$ are the rate coefficients for collional ionization and radiative recombination. The recombination here is what is called case B, which excludes recombination to the ground state. From \cite{Draine_book}, we find 
\begin{equation}
    \alpha_{ci} = 5.466 \times 10^{-9} C T_4^{1/2} e^{-I/k_BT} \, \mathrm{cm^3 \, s^{-1}}
\end{equation}
where $C$ is a constant of order unity and $T_4$ is the temperature in units of $10^4$ K and $I/k_B=157800$ K. 
For case-B recombination, 
\begin{equation}
    \alpha_{rr} = 2.54 \times 10^{-13} \, T_4^{-0.8163 - 0.0208 \ln T_4} \, \mathrm{cm^3 \, s^{-1}}.
\end{equation}
Using these formulas, we find that at $20,000$ K the ionization fraction is $0.95$ and at $8000$ K it is $4.2 \times 10^{-5}$, the enormous range being because of the exponential sensitivity to temperature in the collisional-ionization rate. As mentioned above, at temperatures above $8000$ K the gas cools very rapidly to that temperature; so what is relevant for the inflow into the UDMH is the ionization fraction at $8000$ K. 

\subsection{Compton Drag and Hubble Damping}

The ionization fraction is important, because ionized gas undergoes significant Compton drag \cite{Umemura_Fukue_1994PASJ...46..567U,Umemura_Loeb_Turner_1993}, and 
Compton cooling (see \Eq{Comp_Cool} below). The drag 
can affect both  the rate of accretion,
and the dissipation of angular momentum. The Compton drag is $- \beta v_\infty$, where

{\begin{align}
    \beta =& \frac{4 \sigma_T \epsilon_{\gamma 0} \chi_e}{3 \mu m_p c} (1 + z)^4 \\
    =& 1.8 \times 10^{-15} \frac{\chi_e}{10^{-4}} \left( \frac{1+z}{1100} \right)^4 \, \mathrm{s^{-1}}. 
    \label{beta}
\end{align}
Here $\sigma_T$ is the Thomson cross-section, $\epsilon_{\gamma 0} (= a T^4)$ is the energy density in the cosmic micro-wave background at the present epoch, $\chi_e$ is the ionization fraction of the gas, $\mu$, the proton number per electron, is $0.6$ for primordial gas. 

The gas is moving at a streaming velocity $v_\infty(z) = v_\infty(z_R) (1+z)$ with an rms velocity at recombination $v_{\infty}(z_R) = 30 \, \mathrm{km \, s^{-1}}$ \cite{Tsel_2010}, so that ratio of the Compton drag acceleration to the gravitational acceleration at the accretion radius is 
\begin{align}
    \frac{\beta G M }{v_{\infty}^3} = 8.9 \times 10^{-4} \left( \frac{\chi_e}{10^{-4}} \right) \left( \frac{1+z}{1100}\right)
    \left( \frac{v_{\infty}(z_R)}{ 30 \, \mathrm{km \, s^{-1}}}\right)^{-3}. 
\end{align}
We conclude therefore that, unless there are internal sources of ionization -- a possibility explored in \cite{Umemura_Loeb_Turner_1993} -- the effect of Compton drag is negligible. 

We also have to consider how important is the expansion of the universe in the BHL process. To check that let us compare the Hubble damping $Hv$ with the gravitational acceleration $GM/r_{acc}^2$ at the accretion radius. Assuming that $v \approx v_{eff} \approx v_\infty(z)$, we have 
\begin{equation}
\frac{Hv}{GM/r_{acc}^2} = 0.09 \left[ \frac{M}{10^5 \, M_\odot}\right] \left[ \frac{1100}{1+ z} \right] ^{3/2}
\left( \frac{v_{\infty}(z_R)}{ 30 \, \mathrm{km \, s^{-1}}}\right)^{-3}
\end{equation}
Note that if the streaming velocity were small compared to $c_s$, $r_{acc}$ would be defined by $c_s$, and, consequently, the Hubble damping term would no longer be negligible compared to the gravitational acceleration; in other words, in this regime, the accretion efficiency can be greater when the streaming velocity dominates over the sound speed. 

The Hubble damping can become significant at redshifts below about $220$. But, as we will see below, we are interested only in the accretion for $z > 400$.

\subsection{Accretion Rate and Mass Accumulation}

As reviewed in \cite{Edgar_2004}, the accretion by the BHL process is limited to gas within an accretion radius 
\begin{equation}
    r_{acc} = \frac{2 G M}{v_{eff}^2},
\end{equation}
where $v_{eff}^2 = v_\infty^2 + c_\infty^2$, $v_\infty$ being the streaming velocity at large distances from the UDMH and $c_\infty$ the sound speed there. 

The accretion rate in the BHL process is (\cite{Edgar_2004})
\begin{equation}
    \frac{d M}{d t} = \frac{4 \lambda \pi G^2 M^2 \rho_{\infty}}{v_{eff}^3},
\end{equation}
where $\rho_{\infty}$ is the density of the baryons far away from the UDMH, and $\lambda$ is an efficiency factor that takes into account Hubble expansion and Compton drag. This factor was developed by \cite{Ricotti_2007} in the context of Bondi (spherical) accretion and used by \cite{Ricotti_2008} and \cite{Jangra_etal_2025} in the more general context of BHL accretion. The $\lambda$ factor in equations $3.4$ to $3.6$ of \cite{Jangra_etal_2025} implicitly contain a suppression in the efficiency factor when $v_{eff}$ drops towards $c_s$ (which we mentioned in the last section, when comparing Hubble damping to gravitational acceleration). 
For $\rho_\infty (z_R) = 5.6 \times 10^{-22} \, \mathrm{gm \, cm^{-3}}$, $M = 10^5 \, M_\odot$, and $v_{eff} = 30 \, \mathrm{km \, s^{-1}}$, the standard values we adopt for our calculations below, $d M/d t = 4.6 \lambda \times 10^{22} \, \mathrm{g \, s^{-1}} = 7.3 \lambda \times 10^{-4} \, \mathrm{M_\odot \, yr^{-1}}$. 

In \cite{Jangra_etal_2025} $\lambda$ 
has been calculated for various scenarios for redshifts below $10^4$. For redshifts before recombination, the strong ionization increases the Compton drag and thus decreases the accretion efficiency. On the other hand, at lower redshifts the accretion radius increases and the Hubble expansion factor consequently suppresses accretion. Finally, as the central mass $M$ increases, the accretion radius increases and thus the suppression due to Hubble expansion is more severe. All of this combines to ensure that for $M = 10^5 \, M_\odot$, the largest mass considered by \cite{Jangra_etal_2025}, $\lambda$ remains close to unity down to $z \approx 400$. This agrees with what we find above -- that the effects of Compton drag and Hubble expansion are small in the redshift interval and mass range we are looking at.

As we will see below, this redshift interval is important for another reason, related to the onset of molecular-line cooling. For all of these reasons $M = 10^5 \, M_\odot$ appears to be a sweet spot for the purposes of accretion.

The two speeds that contribute to $v_{eff}$ evolve differently as the universe expands. The streaming velocity simply drops as $(1 + z)$. The sound speed goes as $\sqrt{T}$, where $T$ is the temperature of the gas, which, until $z$ drops to about $200$ 
\cite{BL01}, is determined by the temperature of the CMBR. Thus, in the redshift range that we are interested in
\begin{align}
    v_{eff}^2 &= v_{\infty}^2(z_R) \left( \frac{1+ z}{1100} \right)^2 + c_{\infty}^2(z_R) \left( \frac{1 + z}{1100} \right) \\
    &= v_{\infty}^2(z_R) \left( 1 + \frac{c_{\infty}^2(z_R)}{v_{\infty}^2(z_R) } \frac{1100}{1+z} \right),
\end{align}
where $z_R$ denotes the recombination-era redshift as before. 
Given that typically $c_{\infty}^2(z_R) / v_{\infty}^2(z_R) \approx 1/36$, $v_{eff} \approx v_\infty$ down to the redshift to which this formula is valid, i.e. $z \approx 200$. Thus, the process of accretion of baryons is dominated by streaming -- i.e. it is effectively cold (supersonic) accretion -- throughout the period of accretion in the redshift range $(1100, 400)$. 

In the case we are considering, 
\begin{align}
        r_{acc} =& \frac{2 G M_{rc}}{v_{\infty}^2(z_R)} \left( \frac{1100}{1+z}\right)^2 \\
        =& 1.0 \left[ \frac{M}{10^5 M_\odot} \right] \left[ \frac{v_{\infty}(z_R)}{30 \, \mathrm{km \, s}^{-1}}\right]^{-2} \left[ \frac{1+z }{1100}\right]^{-2} \, \mathrm{pc}
        \label{racc},
\end{align}
where we have incorporated the fact that the streaming velocity $v_{\infty}$ falls off as $(1+z)$ \cite{Tsel_2010}. 

The accretion rate per redshift interval, assuming a constant-mass UDMH, is
\begin{equation}
    \frac{d M}{d z} = - \left[ \frac{4 \lambda \pi G^2 M^2 \, \rho_{\infty}(z_R)}{v_{\infty}^3(z_R)} \right] \, \left[ \frac{\Omega_{m 0}^{-1/2}H_0^{-1}}{(1 + z)^{5/2}} \right].
    \label{dmdz_mfixed}
\end{equation}

We assume that this accretion begins at recombination and ask ourselves how much mass can accumulate $z = 400$. The reason for this cutoff is that at $z=400$, as discussed below, it is possible that the cooling increases significantly as a result of the formation of molecular hydrogen, leading to a sudden change in the Jeans mass and thus to fragmentation. 
Integrating over the redshift interval $(1100, 400)$ we get
\begin{equation}
    M_{acc} = 1200 \,\lambda \left( \frac{M}{10^5 \, M_\odot} \right)^2 \, M_\odot.
\label{Macc1}
\end{equation}

While the baryons are accreting into the UDMH, the dark-matter halo itself is growing as $1/(1+z)$, as shown by Bertschinger \cite{Bert}. This boosts the accretion rate significantly, since it depends on $M^2$. Recalculating, we find that the accretion rate per redshift interval, assuming a UDMH growing as an $1/(1+z)$, increases to
\begin{equation}
    \frac{d M}{d z} = - \left[ \frac{4 \lambda \pi G^2 M_{rc}^2 \, \rho_{\infty}(z_R)}{v_{\infty}^3(z_R)} \right] \, \left[ \frac{\Omega_{m 0}^{-1/2}H_0^{-1}}{(1 + z)^{5/2}} \right] \left[\frac{1100}{1+z} \right]^2,
    \label{dmdz_mvar}
\end{equation}
where $M_{rc}$ is the mass at recombination. (We continue to assume that the mass accreted is negligible compared to the central mass.) This leads to a larger accreted mass
\begin{equation}
    M_{acc} = 5200  \, \lambda \left( \frac{M_{rc}}{10^5 \, M_\odot} \right)^2 \, M_\odot.
\label{Macc2}
\end{equation}

However, the efficiency factor $\lambda$ will begin to decrease from unity due to Hubble expansion in this
case by $z\sim 400$. We also note that if the UDMH were to form only at a later epoch, by say $z=600$, the 
mass accreted is still a significant fraction, $0.58 M_{acc}$ and $0.78 M_{acc}$ of $M_{acc}$ given in
\Eq{Macc1} and \Eq{Macc2} respectively. This is because most of the mass is accreted at later redshifts when the 
available time for accretion is also longer.

In any case, it appears that a significant amount of baryonic matter, $M_g \sim 10^3 M_\odot$,
can accrete into UDMH of mass $M=10^5 M_\odot$ during the redshift interval $(1100, 400)$, in which atomic cooling is likely to dominate. The dominance of atomic cooling, as we will see below (and as argued by \cite{Wenzer+}, \cite{Latif_etal}, \cite{Choi_etal}), can lead to collapse of the baryonic matter with minimal fragmentation, leading, potentially, to the formation of either a super-massive star or, directly or eventually, a black hole. 

\section{Potential Mechanism for Black-Hole Formation}
\label{sec:DCBH}

Once the gas falls into UDMH, it is re-heated through accretion shocks (\cite{Mo_White}) to the virial temperature, which, as shown above in \Eq{Tvir}, is $1.7 \times 10^4$ K for halo mass of $10^5 \, M_\odot$. 
The average gas density is 
\begin{align}
    \rho_b = 2.4 \times 10^{-20} \left( \frac{M}{10^3 \, M_\odot}\right) \left( \frac{r_{vir}}{0.9 \, \mathrm{pc}}\right)^{-3} {\rm g} \ {\rm cm}^{-3}.
\end{align}
This corresponds to a hydrogen density $n_H \sim X\rho_b/m_p 
\sim 1.1 \times 10^4$ cm$^{-3}$, where we have taken Hydrogen abundance of $X=0.75$ and $m_p$ is the proton mass.
The thermal energy density in this gas is thus $\epsilon_{th}= 3n k_B T / 2 = 
7.6  \times 10^{-8} \, \mathrm{ergs \, cm^{-3}}$ for
$n\sim 2n_H$, as for primordial ionized gas \cite{Mo_White}, and $T$ above.

\subsection{Cooling}

\subsubsection{Atomic Cooling}

The gas heated to virial temperature can cool by atomic cooling, whose rate $\Lambda$ can be determined from the density of hydrogen atoms and the calculated curves of $\Lambda/n_H^2$ (e.g. \cite{BL01}). We have 
\begin{align}
    t_{cool} = \frac{\epsilon_{th}}{\Lambda} = 6.2 \times 10^6 \, \mathrm{s} < 1 \, \mathrm{yr}.
\end{align}
for $\Lambda/n_H^2= 10^{-22} \, \mathrm{ergs \, cm^3 \, s^{-1}}$. In other words, the cooling is 
almost instantaneous compared to the free-fall time,
\begin{align}
    t_{ff} \approx & \sqrt{ \frac{r_{vir}^3}{GM}}   \\
        =& 4.0\times 10^4 \, \mathrm{yr} \left[ \frac{r_{vir}}{0.9 \, \mathrm{pc}} \right]^{3/2} \left[ \frac{M}{10^5 \, M_\odot} \right]^{-1/2}.
\end{align}
However, because $\Lambda/n_H^2$ drops by five orders of magnitude by the time the temperature falls to $8000 $ K, the cooling time-scale rises rapidly as the temperature falls, and, at some point, becomes equal to the free-fall time $t_{ff}$.
At that point where $t_{cool} \approx t_{ff}$, we expect the temperature to stabilize, since, as the gas contracts, there will be a certain amount of adiabatic heating due to the 
work done by pressure during the gas compression.
The temperature of the final configuration is difficult to predict exactly, since the cooling curve drops so rapidly, but one expects
it to be fairly close to $8000$ K.
As this temperature is below the virial temperature of the halo, the gas will 
continue its collapse in the UDMH potential well.

\subsubsection{Role of Compton Cooling}

An ionized gas can also lose energy through the inverse Compton process involving interactions between free electrons and cosmic background radiation. The time-scale for Compton cooling is related to the time-scale for Compton drag (see \Eq{beta}) by 
\begin{align}
    t_{cool} &= \frac{m_e}{m_p}t_{drag} = \frac{m_e}{m_p} \frac{1}{\beta} \nonumber \\
             &\approx 10^4 \left( \frac{10^{-4}}{\chi_e} \right) \left(\frac{1100}{1 + z} \right)^4 \, \mathrm{yr}.
\label{Comp_Cool}
\end{align}
The Compton cooling time-scale, unlike that for atomic cooling, is independent of density and temperature 
(as long as the temperature
is much larger than that of the CMB). It is effective when the ionization fraction is large, but the consequent cooling leads to recombination and a fall in the ionization fraction; thus it is automatically regulated. In the conditions we are considering, this time-scale is much longer than that of atomic cooling, especially as the redshift drops. For example, by $z = 400$, it increases by a factor of $57$. Thus we can neglect Compton cooling compared to atomic cooling.

\subsubsection{Role of Molecular Cooling}

So far we have been assuming that the gas temperature is limited by atomic cooling. However, 
the standard star formation, which of course happens readily in later epochs, when gas clouds are rich in metals and dust, can in principle happen at early epochs as well if an appropriate cooling mechanism is available to bring the temperature down, thus reducing the Jeans mass to star-sized clumps. In the absence of metals, the main mechanism we need to consider is cooling due to the $H_2$ molecule. 
Ito and Omukai \cite{Ito_Omukai_2024}, in their study on the formation of Pop III stars in early epochs examine various pathways for the formation of $H_2$, and conclude that the standard pathway, involving the $H^-$ ion, becomes unavailable at $z > 130 $ (due to the photo-detachment of the $H^{-}$ ion by the CMB). They also study a second pathway, involving the $H_2^+$ ion. This pathway is two orders of magnitude less efficient  than the standard pathway, and therefore negligible when both are available, but it too becomes unavailable at $z > 400$. Thus, 
molecular hydrogen formation is strongly suppressed in the gas collapsing in the UDMH at
redshifts higher than 400 
and there will be no further cooling due to $H_2$. 
In addition, for $z>400$, the CMB temperature $T_{CMB} > 1100$ K is large enough to
keep the low lying rotational levels of any existing $H_2$ in radiative equilibrium, also suppressing
$H_2$ cooling below $T_{CMB}$ \cite{Omukai+2005}.
For a significant amount of gas to be accreted before $z = 400$, the process must begin well before that, as it does in the scenario we have outlined. 

\subsection{Baryon collapse}

Thus, in an atomic cooling UDMH, the gas is essentially in free fall after it falls into the central potential, and maintains a temperature of about $8000$ K. 
The important question that arises at this point is whether the gas fragments.
Since the gas temperature is maintained at $8000$ K, 
the Jeans mass $M_J \propto 1/\sqrt{n}$, and at some point the density will be large enough for fragmentation to begin. 
However, since both the fragmentation and the
free-fall times go as $1/\sqrt{G \rho}$ the fragments may themselves fall in towards the centre of the UDMH potential at the same rate as they individually collapse, so that they eventually coalesce before forming stars from the clumping.
Moreover, for fragmentation to be followed by collapse of individual clumps, the self-gravity of the clumps would have to overcome a comparable tidal force of the cloud as a whole.}
Simulations by \cite{Choi_etal} show that fragmentation is suppressed, and those by \cite{Latif_etal} show that, in the absence of cooling by the $H_2$ molecule, 
even when fragmentation occurs, the fragments indeed tend to coalesce. 

Haiman \cite{Haiman_2012} points out that if the density becomes large enough, the gas may become optically thick to the line radiation that would cool it when optically thin, and that as a result radiation pressure may support 
fragmenting structures against gravity. 
Indeed, this is obtained in simulations (\cite{Bromm_Loeb_2003}) in which, at a temperature of $\approx 10^4$ K, up to $10^6 \, M_\odot$ of gas continues to condense without showing any fragmentation (see also \cite{Choi_etal}). Higher-resolution simulations with the initial conditions envisaged here are needed to shed more light on what happens at small scales.

\subsubsection{Role of angular momentum}

We have assumed that the angular momentum of the infalling gas is negligible, and that it remains negligible even within the virial radius. In standard structure formation models, angular momentum of both baryons and dark matter 
arises due to tidal gravitational torques as a putative dark matter halo expands, turns around
and collapses (cf. \cite{Mo_White}). However in our context, the UDMH already collapses by recombination, when the
baryons are relatively smooth and have not yet responded greatly to gravity (see \Sec{sec:baryon1}). Thus the 
baryons are not expected to acquire angular momentum at this stage, but only later
while they accrete by the BHL process on to the collapsed UDMH.
The process of standard BHL accretion itself is cylindrically symmetric, and ought not to generate much angular momentum. A potential final source of angular momentum in the gas is that due to BHL accretion on to a non-spherical halo. 
However, we envisage the UDMH which collapse by recombination to be from rare $5\sigma-6\sigma$ density peaks, whose
initial mean ellipticities are expected to be $ <0.1$ \cite{Mo_White,Delos_Silk}. Then the UDMH which
forms from their collapse would also be expected to be nearly spherical. Moreover, 
much of the non-radial velocity component will be canceled in the shock, 
where streams from opposite sides of the accreting UDMH meet.

Nevertheless, some angular momentum $L$ could be generated, which we can quantify 
in the usual manner by the dimensionless parameter $\lambda_L =\omega/\omega_0 \sim L/M_gv_cr$, 
the ratio of an angular velocity $\omega=L/M_gr^2$ , associated with the gas of mass $M_g$ and radius $r$ 
by the angular velocity if the gas is supported by rotation $\omega_0 \sim v_c/r$, where $v_c$ is the halo
circular velocity. Assuming $L$, $M_g$ and $v_c$ not to change during the gas collapse, $\lambda_L \propto 1/r$.
Since $\lambda_L\sim 1$ for a rotationally supported system, we expect the gas to collapse
to a final radius $r_f \sim \lambda_{Li} r_i$, where $r_i$ is the initial radius and $\Lambda_{Li}$ 
the initial value of $\lambda_L$. For $\lambda_{Li} \sim 0.01-0.1$, the gas will collapse
to a compact thick disk of radius $r \sim 0.1r_{vir} - 0,01 r_{vir}$, partially supported by angular momentum,
before the final collapse. 
Along the direction of the angular momentum 
the gas can collapse further until supported by thermal pressure, to a thickness $\cal{H}$,
as estimated below.

\subsubsection{Angular momentum transport and accretion}

One process that could lead to a loss of angular momentum, as pointed out in \cite{Loeb1993}, 
is Compton drag (see \Eq{beta}). The time-scale corresponding to this drag is 
\begin{equation}
    t_{drag} = \frac{1}{\beta} = 1.8 \times 10^7 \left( \frac{10^{-4}}{\chi_e}\right) \left( \frac{1100}{1+z}\right)^{4} \, \mathrm{yr}.
\end{equation}
For $t_{drag}$ to be comparable to $t_{ff}$, the ionization fraction would have to be about $0.1$, and even 
larger at lower redshifts. This is realistically possible only if there is a source of ionization as envisaged by \cite{Loeb1993}.

Whatever residual angular momentum there is will cause the baryons to form a disk. From the arguments given above we expect the radius of the disk to be $R \sim r_f \sim \lambda_{Li}r_{vir}$. The thickness of the disk is expected to be 
$\mathcal{H} = c_s/\Omega = R c_s/v_c$ \cite{Frank_etal_book}. 
Here $v_c$ is the circular velocity as determined by the halo potential and given in \Eq{vc}, while
$\Omega = v_c/r$ is the rotation frequency of the gas. 
Viscous forces can then lead to a slow 
loss of angular momentum and accretion into the centre of the potential. The accretion time-scale from a disc of size $R$, for a kinematic viscosity $\nu$ is $t_{acc} \sim R^2/\nu$ \cite{Frank_etal_book}. Using the $\alpha$ prescription \cite{Frank_etal_book}, we can write $\nu = \alpha c_s \mathcal{H} $. Thus
\begin{equation}
    t_{acc} \approx \frac{1}{\alpha \Omega} \left( \frac{v_c}  {c_s}\right)^2 = \left( \frac{\lambda_{Li}}{\alpha} \right) \left( \frac{r_{vir} v_c}{c_s^2}\right).
\end{equation}
For $\lambda_{Li} = 0.01$, $\alpha=0.01$, and $c_s = 10 \, \mathrm{km \, s^{-1}}$ (for an $8000$ K gas with mean molecular weight $\mu = 1$) and the other quantities with the values assumed above, we get $t_{acc} = 1.9 \times 10^5$ yr, which is about $10 \%$ of the age of the universe at $z = 400$. 
The value of $\alpha$ is governed by the level of turbulence. This could originate from 
the accretion shock to begin with and amplified by the initial collapse \cite{Hennebelle}. The turbulence 
in a collapsing system can also lead to super-exponential dynamo amplification of magnetic fields \cite{Irshad+25} and both will
lead to angular momentum transport. Turbulence and magnetic fields  can additionally arise due to the
magneto-rotational instability (MRI) \cite{BH_review}. In the context of a weakly ionized disk, MRI is less efficient,
although \cite{Flock+12} find $\alpha\sim 0.01$ can obtain even in weakly ionized disks given a sufficiently
electrically conducting plasma.

More efficient transport of angular momentum and accretion can obtain if the rotating 
disk becomes self-gravitationally unstable.
To check whether such a disk is stable, we use the Toomre stability parameter
\begin{equation}
    Q = \frac{c_s \kappa}{\pi G \Sigma},
\end{equation}
where $\kappa$ is the epicyclic frequency, which is $\sqrt{2} \Omega$ for a flat rotation curve. If all the gas forms a disk, then we have, assuming a disk of uniform thickness,
\begin{equation}
    \rho = \frac{M_g}{\pi R^2 \mathcal{H}} = \frac{M_g}{\pi \lambda_{Li}^2 r_{vir}^2 \mathcal{H}}.
\end{equation}
Using $\Sigma = \rho \mathcal{H}$, we get 
\begin{equation}
            Q = \frac{\sqrt{2}\lambda_{Li}v_c c_s r_{vir}}{G M_g}.
\end{equation}
For $\lambda_{Li} = 0.01$, $v_c = 22$ km s$^{-1}$, $c_s = 10$ km s$^{-1}$, $r_{vir} = 0.9 $ pc, and $M_g = 10^3 \, M_\odot$, we get the $Q = 0.63$, which suggests that it is unstable ($Q > 1$ for stability). Note that for $\lambda_{Li} = 0.01$, the gas mass contained within the boundary of the disk is comparable to the halo mass within the same radius, assuming the density distribution of the UDMH is close to an isothermal sphere.
This means that the disk would be almost self-gravitating. 
Such self-gravitating, Toomre unstable disks efficiently transport angular momentum outwards
leading to mass accretion, with an effective $\alpha \sim 0.01-1$ (see \cite{GI_Review} for a review
in the context of circum-stellar disks). The resulting accretion can occur then even as fast as a few rotation time-scales.
More detailed calculations, taking the time dependence of the 
BHL accretion process, the growth of the UDMH mass and the subsequent build up of the disk 
mass, are needed to firm up the above picture.
\footnote{
Interestingly, with $Q \propto v_c r_{vir}/M_g$, and as 
$M \propto (1+z)^{-1}$, 
$v_c r_{vir} \propto M^{2/3} (1+z)^{-1/2} \propto (1+z)^{-7/6}$, 
while the accreted gas mass, using Eq.~(\ref{Macc2}), increases as $M_g \propto (1+z)^{-7/2}$.
This can result in 
a possible decrease in $Q\propto (1+z)^{7/3}$.
}

Overall, it appears that about $10^3 M_\odot$ of gas can be accreted due to the BHL process 
before $z\sim 400$, by a $10^5 M_\odot$ UDMH forming by the epoch of recombination,  
undergo atomic cooling and collapse significantly to the UDMH center. Fragmentation into smaller masses 
is prevented as $H_2$ molecule formation and the resulting molecular cooling is strongly suppressed at such high redshifts due to the influence of the 
CMB. The angular momentum acquired by the gas is expected to be small enough, at least in some halos,
that the resulting gas disk is compact and self-gravitationally (Toomre) unstable.
Both the resulting gravitational torques and viscous forces will cause
rapid accretion to form a compact supermassive star or potentially lead
to a direct collapse blackhole. 
The possibility of such direct-collapse black holes, in the 
stages of galaxy formation were first considered in \cite{Bromm_etal}, \cite{Begelman_etal}, and \cite{Lodato_etal}.

\section{\label{sec:discussion}Discussion and conclusions}

We have proposed a scenario for the formation of black holes by the accretion of baryons between recombination and $z =400$ into ultra-dense dark-matter halos (UDMH). 
UDMH can form in abundance from more typical density
fluctuations in models where rarer large-amplitude curvature fluctuations collapse 
to form primordial black holes (see for example \cite{Delos_Silk}).
Using more conservative initial conditions, and with 
amplitudes of $\zeta$ on small-scales significantly below current cosmic microwave background distortion limits, we show that UDMH of mass $\sim 10^5 M_\odot$ can form by very high redshifts post recombination. We estimate the abundance of 
such UDMH, collapsing at such high redshifts from the rare $5\sigma$ to $6\sigma$ density
fluctuations to be comparable to that of galaxies. 

The accretion of baryons into such UDMH happens through the Bondi-Hoyle-Lyttleton (BHL) process, because of large-scale streaming of baryons with respect to the dark matter in the post-recombination era. We find that a number of factors combine to make this process efficient between $z=1100$ and $z=400$. First, when the streaming is at several times the sound speed, 
which it can be at these red-shifts, and the UDMH mass is of order $10^5 \, M_\odot$, the accretion radius is such that Hubble damping is unimportant. Second, the shocks formed in the BHL process heat the gas, in the beginning, to temperatures of order 
$2\mu \times 10^4$ K and, this gas is cooled almost instantaneously by atomic cooling to a temperature of about $8000$ K, at which the ionization fraction in the ambient conditions is typically less than $10^{-4}$; later in the process, the post-shock temperature is lower than $8000$ K, and the ionization fraction remains low; the gas falling into the UDMH is thus largely neutral, making Compton drag negligible, and thus increasing the efficiency of accretion. 

Once the gas falls into the virialized UDMH, it gets reheated to the virial temperature of $\sim 2 \times 10^4$ K  before being cooled. Once again, because of efficeint atomic cooling the temperature of the gas drops to $8000$ K as it sinks into the UDMH 
core over the free-fall time-scale, which we estimate to be a few times $10^4$ years. As the gas sinks into the UDMH there is a certain amount adiabatic heating as potential energy is converted into kinetic energy. We expect the cooling and heating to balance when the cooling time becomes comparable to the free-fall time. 
Importantly, at redshifts greater than $z = 400$ the cooling due the $H_2$ molecule is unimportant as its formation itself is
suppressed due the influence of the CMB. Thus we expect the gas to maintain a temperature of $\sim 8000$ K as it becomes more compact. Since this temperature is less than the virial temperature, gas continues to collapse inward. 
It has been argued by several authors \cite{Wenzer+,Latif_etal,Choi_etal}, that under these conditions, 
where atomic cooling dominates, and molecular cooling is suppressed, 
 collapse of the baryonic matter occurs with minimal fragmentation.
 We also note that collapse of individual clumps following 
 fragmentation is unlikely as it proceeds on the same time-scale as the overall 
 collapse, and has to overcome tidal forces. 

What happens to the gas as it sinks deeper into the UDMH depends crucially on angular momentum. We argue that,   
since the accretion process begins well after the UDMH has collapsed, and given the gas
is almost homogeneous at recombination, the BHL accretion process itself is cylindrically symmetric. Moreover, as rare halos tend to be more spherically symmetric, there is a good chance that in some cases the angular momentum of the collapsing gas will be close to zero, leading to direct collapse. We also show that if the gas has a small 
enough angular momentum to form a disk that is $0.01$ times the virial radius, 
viscous forces, even with an viscosity parameter $\alpha\sim 0.01$ can cause accretion on a time-scale of order
$2\times 10^5$ yr, which is $10\%$ of the age of the universe at $z=400$. Additionally,
the resulting disk is likely to be gravitationally (Toomre) unstable. Such an unstable disk can 
also transport angular momentum outward even more efficiently.

There are a number of caveats to our estimate for the abundance of 
$10^3 M_\odot$ DCBH, which are worth mentioning. Our aim was to illustrate that,
in principle, their abundance can be large enough, that a fraction of galactic super massive 
black holes could grow from such seed DCBH. 
However, the abundance of UDMH which lead to DCBH as we envisage, 
will be smaller if the level of curvature fluctuations is smaller. 
For example, if only curvature fluctuations greater than $6.5\sigma$ collapse 
to UDMH early enough to form DCBH, 
their abundance is $\sim 2.6 \times 10^{-5}$ Mpc$^{-3}$,
compared to $N \sim 6.3 \times 10^{-4}$ Mpc$^{-3}$ for $6\sigma$ fluctuations 
(adopting Gaussian probability distribution). This is much lower than galactic abundance, 
but comparable to the abundance of
high redshift Little Red Dots (LRDs) which are believed to host accreting 
black holes \cite{Kocevski2025}.
Moreover, if the accreting baryons acquire angular momentum, such that $\lambda_{Li}$ is
larger than $0.01$ that we adopt, the accretion disk will not become Toomre unstable and the 
process to form a compact object will become less efficient. 
Finally, the growth to become a SMBH hosting an AGN also depends on 
subsequent accretion history, and also  the duty cycle 
of AGN activity.

One may also wonder if the abundance of DCBH that we estimate violates any
observational constraints. Carr et al \cite{Carr2026}
summarize constraints on the the fraction $f_{BH}$ of the dark matter in primordial 
black holes (PBH) of various masses, and these limits can be a guide to constraints on the
DCBH as well. 
For black holes with mass $M_{BH} = 10^3 M_\odot$, they point out that the strongest constraints are 
from works on updated CMB bounds, which give $f_{BH} < 10^{-8}$
(see Fig 4 in \cite{Carr2026} and also \cite{Serpico2020}). 
For the range of abundances of $10^3 M_\odot$ DCBH $N \sim 10^{-3}-10^{-5}$ Mpc$^{-3}$,
we have $f_{BH} = N M_{BH}/\rho_{d0} \sim 3 \times 10^{-11} - 3 \times 10^{-13}$, much 
smaller than this upper limit.

There are a number of ideas in this paper that merit further exploration. The ionization fraction of the gas plays a crucial role because it determines the strength of the Compton damping, which can be crucial in two respects: if the ionization fraction is high in the initial stages of accretion, the efficiency falls; on the other hand, if it is high in the later stages, once the gas has settled into the halo, then it can damp out angular momentum very effectively (\cite{Umemura_Loeb_Turner_1993}). 
Maintaining a high ionization fraction would require external sources of radiation, which we do not consider here.
The precise role of cooling by the $H_2$ molecule is important because it is related to the temperature of gas in the halo, which in turn determines the Jean mass and thus the likelihood of fragmentation. Another 
important factor is angular momentum and 
the potential formation of an accretion disk. It will be useful to capture these features in simulations
of the BHL accretion into UDMH. Equally, the observational consequences of such UDMH formation 
and their hierarchical clustering and growth at later redshifts will be
important to explore further.
The DCBH formed in some of these halos can also feed back on the formation of the
first stars and galaxies \cite{Zhang}. Such effects
will also potentially constrain the power spectrum on small-scales.

There appears to be a sweet spot in UDMH mass of about $10^5 \, M\odot$ and a window between $z = 1100$ and $z = 400$ when baryon accretion could be efficient enough for $\sim 10^3 \, M_\odot$ to be accumulated in the UDMH 
and collapse efficiently to form a compact supermassive star or a direct-collapse black hole. 

\begin{acknowledgments}
We thank Dipankar Bhattacharya, R. Srianand and S. Sridhar for useful discussions.
BP thanks the Inter-University Centre for Astronomy and Astrophysics 
for its hospitality during several visits there.
KS was partially supported by the Alexander Von Humboldt Foundation through the
Carl Friedrich von Siemens Research Award. He thanks Volker Springel his host, 
for warm hospitality at the Max-Planck Institute for Astrophysics, Garching.
We thank the referee for several very constructive comments. 
\end{acknowledgments}

\appendix

\section{\label{sec:baryon1}Baryon response before recombination}

In the radiation dominated universe, baryons are strongly coupled to
photons and oscillate as sound waves. However these oscillations are damped
by radiative viscosity (Silk damping), and for the small mass scales
of interest in this paper, the baryon fluctuation $\delta_b$ is driven to zero initially.
However the photon mean free path $l_\gamma = 1/n_e \sigma_T$ (with the electron number $n_e \propto 1/a^3$ 
and $\sigma_T$ the Thompson cross section), increases with time as $a^3$. When it becomes large enough, such that $(k/a)l_\gamma >1$, 
then the baryons are no longer tightly coupled to photons. They can begin to respond to the growing dark matter perturbations, although still damped
by free-stream radiative viscosity. Moreover, baryons on larger scales are still part of the
acoustic oscillations and so on small scales of interest, one also has to
take account of the free-steaming velocity ${\bf v}_{bc}$ between the baryons and the dark matter \cite{Tsel_2010}. 
This has a Gaussian probability distribution with a variance $v_{bc} \sim 30$ km s$^{-1}$ at recombination
and will also suppress baryons falling into the dark matter potential wells.

It is of interest to nevertheless ask how
much growth of $\delta_b$ is possible at this stage.
The evolution equations for baryon density perturbation $\delta_b$ and peculiar velocity ${\bf u}_b$ in 
Fourier space, 
\footnote{The Fourier space variables are defined with a 'hat'; for example $\hat{\delta}_b$
is the Fourier transform of $\delta_b$.}
are \cite{Tsel_2010}
\begin{equation}
\frac{\partial\hat{\delta}_b}{\partial t} +\frac{1}{a} i {\bf k}\cdot{\bf v_{bc}}\hat{\delta}_b  = - \frac{1}{a} i{\bf k}\cdot\hat{\bf u}_b,
\label{mass}
\end{equation}
\begin{equation}
   \frac{\partial\hat{\bf u}_b}{\partial t} +\left( \frac{i {\bf k}\cdot{\bf v_{bc}}}{a}
   + H+\beta \right)\hat{\bf u}_b = 
   -\frac{1}{a} i{\bf k}\hat{\delta}_b c_s^2
   -\frac{1}{a} i {\bf k} \hat{\phi} .
   \label{momentum}
\end{equation}
Here we have taken the baryon pressure perturbation to be $\rho_b \hat{\delta}_b c_s^2$, with $\rho_b$ the mean baryon density and
$c_s$ the sound speed. The potential perturbation $\hat{\phi}$ satisfies the Poisson equation with the total matter density
perturbation $4\pi G\rho_m \hat{\delta}_m $
as the source,
\begin{equation}
    -\frac{k^2}{a^2} \hat{\phi} = 4\pi G \rho_m \hat{\delta}_m 
    = \frac{3}{2} H^2 \hat{\delta}_m,
    \label{potential}
\end{equation}
where the second equality holds at high redshifts, where dark energy contribution is
negligible.
The momentum equation \Eq{momentum} incorporates not only 
Hubble damping term $H\hat{\bf u}_b$ but also  
the Compton drag on the baryons ($-\beta \hat{\bf u}_b$), due to 
free streaming photons on scales $k l_\gamma/a > 1$, where 
$\beta$ is given in \Eq{beta} with $\chi_e=1$ before recombination.

Let us compare the Compton drag with the Hubble damping, $H\hat{\bf u}_g$, where
$ H = H_0\sqrt{\Omega_m} (1+z)^{3/2}$, and the damping due to baryon-dark matter
streaming, $(k v_{bc}/a) \hat{\bf u}_g$. We have
\begin{align*}
\frac{H}{\beta} &\approx 2.8 \times 10^{-3} \chi_e^{-1} \left( \frac{1+z}{1100} \right)^{-5/2} \\
\frac{k v_{bc}}{a\beta} &\approx 6 \times 10^{-3} \left(\frac{v_{bc}}{30 \mathrm{km} \mathrm{s}^{-1}}\right) 
\left(\frac{k}{270 \mathrm{Mpc}^{-1}}\right) \chi_e^{-1} \\
 & \quad \times \left( \frac{1+z}{1100} \right)^{-3}
\end{align*}
Thus both these damping terms can be neglected compared
to the Compton damping in Eq.~(\ref{momentum}). 
The ratio 
of the acceleration $\partial\hat{\bf u}_b/\partial t$ to the Compton drag 
term is also
of order $H/\beta \ll 1$ and so can be neglected. 
Moreover, using \Eq{momentum} and \Eq{potential} the ratio of the restoring force due to baryonic pressure gradient 
compared to the gravitational driving  is given by,
\[
\frac{c_s^2\hat{\delta}_b}{\hat{\phi}}
\approx 0.3 \ \frac{\hat{\delta}_b}{\hat{\delta}_m} \left(\frac{k}{270 \mathrm{Mpc}^{-1}}\right)^2.
\]
Here we have taken $h=0.7$, assumed $c_s^2 \propto 1/a$ since the Baryon temperature is
locked to the CMB temperature before recombination, and $c_s=6$ km s$^{-1}$ at recombination.
Thus the baryon pressure gradient can also be neglected in \Eq{momentum} compared to
the gravitational driving, if 
$\hat{\delta}_b\ll \hat{\delta}_m$ before recombination. We show this inequality indeed holds
in a self-consistent calculation which neglects the baryon pressure perturbation
to begin with.

We then have $\hat{\bf u}_b =  -i {\bf k} \hat{\phi}/(\beta a)$ and substituting this in
\Eq{mass} and integrating, we have
\begin{equation}
    \hat{\delta}_b(t) = \int_{t_i}^t dt' 
       e^{-i\Phi(t,t')} \ \frac{3}{2} \frac{H^2 \hat{\delta}_m}{\beta}
\label{deltab}
\end{equation}
where
\[
\Phi(t,t')= \int_{t'}^{t} dt'' \ \frac{{\bf k}\cdot{\bf v}_{bc}}{a} 
\]
is a time varying phase factor due to the streaming velocity which damps the growth of $\hat{\delta}_b$.

Suppose we first consider the maximum value of $\hat{\delta}_b$ that can arise by recombination
in regions of the universe where the streaming velocity is small enough. 
We adopt $\beta$ from \Eq{beta}, $H^2 = H_0^2 \Omega_m (1+z)^3$, $\hat{\delta}_m = f_d \hat{\delta}_d = f_d \hat{\delta}(z_R)/(1+z)$,
where $f_d=\Omega_d/(\Omega_d+\Omega_b)$ is the dark matter fraction and $\hat{\delta}_d(z_R)$ is the
dark matter density contrast at recombination. Then integrating \Eq{deltab} from
redshift $z_i < z_R$, we find
\begin{equation}
    \hat{\delta}_b(z_R) \approx 1.1 \times 10^{-3} \hat{\delta}_d(z_R) 
    \left(1 - \left(\frac{1100}{1+z_i}\right)^{7/2} \right),
    \label{deltaR}
\end{equation}
where we have taken $1+z_R=1100$.
We see that $\hat{\delta}_b/\hat{\delta}_d \ll 1$, even in regions where 
streaming motions have negligible effect.

To obtain an order of magnitude estimate of the effect of streaming, we 
assume a constant streaming velocity, a matter dominated universe and 
integrate from some initial redshift $z'$ to redshift $z$.
We then get 
for the phase factor,
\begin{equation}
\Phi(z,z')\approx \frac{2{\bf k}\cdot{\bf v}_{bc}}{H_0 \sqrt{\Omega_m
(1+z)}} \left(1 - \frac{\sqrt{1+z}}{\sqrt{1+z'}} \right) 
\label{Phi}
\end{equation}
For $k \sim 270$ Mpc$^{-1}$, a streaming velocity with magnitude $v_{bc} 
\sim 30$ km s$^{-1}$ and by the recombination redshift $z=1100$, 
we estimate the pre factor in \Eq{Phi} to be $\sim 12.7$.
Thus there is a strong cancellation effect in the integral over time
in \Eq{deltab}, due to the rapidly varying phase factor $e^{-i\Phi}$, except for
$t'$ (or $z'$) close to the recombination epoch $t_R$ (or $z_R$). For the above parameters, 
we can estimate that only $z'$ for which 
$\sqrt{1+z_R}/\sqrt{1+z'} < 1 -1/12.7 \sim 0.92$ contribute to the integral in \Eq{deltab}.
Using this in \Eq{deltaR}, we estimate that $\hat{\delta}_b(z_R)$ is reduced by a further
factor $\sim 0.4$. Thus we expect that the baryons develop very small density contrast
by recombination, with $\delta_b \sim 0.4 \times 10^{-3}$, 
even if the dark matter has become nonlinear with $\delta_d \sim 1$. 
Within the virialized
regions of collapsed halos, the dark matter overdensity is much larger, by factor of 
order $200$ or larger, and then one may expect $\delta_b \sim 0.1$ in the halo
interiors, the exact value depending on when the UDMH virialized. 
The major reason for these small values of $\delta_b$ is the strong Compton drag exerted by 
the radiation background. After recombination this drag is drastically 
reduced, and \Sec{sec:baryon2} considers this post recombination epoch.

\bibliography{BC}

@misc{fei2026,
      title={Direct pathway to the Early Supermassive Black Holes: A Red Super-Eddington Quasar in a Massive Starburst Host at $z=7.2$}, 
      author={Qinyue Fei and Seiji Fujimoto and Gabriel Brammer and Ruancun Li and Luis C. Ho and Volker Bromm and Javier Álvarez-Márquez and Yoshihisa Asada and Guillermo Barro and Luis Colina and Pratika Dayal and Steven L. Finkelstein and Johan P. U. Fynbo and Michele Ginolfi and Kohei Inayoshi and Vasily Kokorev and Gene C. K. Leung and Jorryt Matthee and Romain A. Meyer and Rohan P. Naidu and Masafusa Onoue and Pablo G. Pérez-González and Charles L. Steinhardt and Francesco Valentino and Fabian Walter and Mengyuan Xiao and Haowen Zhang},
      year={2026},
      eprint={2602.12325},
      archivePrefix={arXiv},
      primaryClass={astro-ph.GA},
      url={https://arxiv.org/abs/2602.12325}, 
}

@ARTICLE{Paccuci23,
       author = {{Pacucci}, Fabio and {Nguyen}, Bao and {Carniani}, Stefano and {Maiolino}, Roberto and {Fan}, Xiaohui},
        title = "{JWST CEERS and JADES Active Galaxies at z = 4-7 Violate the Local M $_{{\textbullet}}$-M $_{{\ensuremath{\star}}}$ Relation at >3{\ensuremath{\sigma}}: Implications for Low-mass Black Holes and Seeding Models}",
      journal = {\apjl},
         year = 2023,
        month = nov,
       volume = {957},
       number = {1},
          eid = {L3},
        pages = {L3},
          doi = {10.3847/2041-8213/ad0158},
archivePrefix = {arXiv},
       eprint = {2308.12331},
 primaryClass = {astro-ph.GA},
       adsurl = {https://ui.adsabs.harvard.edu/abs/2023ApJ...957L...3P}
}

@article{Paccuci15,
    author = {Pacucci, Fabio and Volonteri, Marta and Ferrara, Andrea},
    title = {The growth efficiency of high-redshift black holes},
    journal = {Monthly Notices of the Royal Astronomical Society},
    volume = {452},
    number = {2},
    pages = {1922-1933},
    year = {2015},
    month = {07},
    issn = {0035-8711},
    doi = {10.1093/mnras/stv1465},
    url = {https://doi.org/10.1093/mnras/stv1465},
    eprint = {https://academic.oup.com/mnras/article-pdf/452/2/1922/18508837/stv1465.pdf},
}

@article{Jeon25b,
doi = {10.3847/1538-4357/ade2e1},
url = {https://doi.org/10.3847/1538-4357/ade2e1},
year = {2025},
month = {jul},
publisher = {The American Astronomical Society},
volume = {988},
number = {1},
pages = {110},
author = {Jeon, Junehyoung and Liu, Boyuan and Taylor, Anthony J. and Kokorev, Vasily and Chisholm, John and Kocevski, Dale D. and Finkelstein, Steven L. and Bromm, Volker},
title = {The Emerging Black Hole Mass Function in the High-redshift Universe},
journal = {The Astrophysical Journal},
}

@ARTICLE{Jeon25a,
       author = {{Jeon}, Junehyoung and {Bromm}, Volker and {Liu}, Boyuan and {Finkelstein}, Steven L.},
        title = "{Physical Pathways for JWST-observed Supermassive Black Holes in the Early Universe}",
      journal = {\apj},
         year = 2025,
        month = feb,
       volume = {979},
       number = {2},
          eid = {127},
        pages = {127},
          doi = {10.3847/1538-4357/ad9f3a},
archivePrefix = {arXiv},
       eprint = {2402.18773},
 primaryClass = {astro-ph.GA},
       adsurl = {https://ui.adsabs.harvard.edu/abs/2025ApJ...979..127J}
}

@ARTICLE{Priya24,
       author = {{Natarajan}, Priyamvada and {Pacucci}, Fabio and {Ricarte}, Angelo and {Bogd{\'a}n}, {\'A}kos and {Goulding}, Andy D. and {Cappelluti}, Nico},
        title = "{First Detection of an Overmassive Black Hole Galaxy UHZ1: Evidence for Heavy Black Hole Seed Formation from Direct Collapse}",
      journal = {\apjl},
         year = 2024,
        month = jan,
       volume = {960},
       number = {1},
          eid = {L1},
        pages = {L1},
          doi = {10.3847/2041-8213/ad0e76},
archivePrefix = {arXiv},
       eprint = {2308.02654},
 primaryClass = {astro-ph.HE},
       adsurl = {https://ui.adsabs.harvard.edu/abs/2024ApJ...960L...1N}
}

@ARTICLE{Trinca2023,
       author = {{Trinca}, Alessandro and {Schneider}, Raffaella and {Maiolino}, Roberto and {Valiante}, Rosa and {Graziani}, Luca and {Volonteri}, Marta},
        title = "{Seeking the growth of the first black hole seeds with JWST}",
      journal = {\mnras},
         year = 2023,
        month = mar,
       volume = {519},
       number = {3},
        pages = {4753-4764},
          doi = {10.1093/mnras/stac3768},
archivePrefix = {arXiv},
       eprint = {2211.01389},
 primaryClass = {astro-ph.GA},
       adsurl = {https://ui.adsabs.harvard.edu/abs/2023MNRAS.519.4753T}
}

@ARTICLE{Zhang,
       author = {{Zhang}, Saiyang and {Liu}, Boyuan and {Bromm}, Volker and {Jeon}, Junehyoung and {Boylan-Kolchin}, Michael and {K{\"u}hnel}, Florian},
        title = "{How do Massive Primordial Black Holes Impact the Formation of the First Stars and Galaxies?}",
      journal = {\apj},
         year = 2025,
        month = jul,
       volume = {987},
       number = {2},
          eid = {185},
        pages = {185},
          doi = {10.3847/1538-4357/adddb4},
archivePrefix = {arXiv},
       eprint = {2503.17585},
 primaryClass = {astro-ph.GA},
       adsurl = {https://ui.adsabs.harvard.edu/abs/2025ApJ...987..185Z}
}

@ARTICLE{Serpico2020,
       author = {{Serpico}, Pasquale D. and {Poulin}, Vivian and {Inman}, Derek and {Kohri}, Kazunori},
        title = "{Cosmic microwave background bounds on primordial black holes including dark matter halo accretion}",
      journal = {Physical Review Research},
         year = 2020,
        month = may,
       volume = {2},
       number = {2},
          eid = {023204},
        pages = {023204},
          doi = {10.1103/PhysRevResearch.2.023204},
archivePrefix = {arXiv},
       eprint = {2002.10771},
 primaryClass = {astro-ph.CO},
       adsurl = {https://ui.adsabs.harvard.edu/abs/2020PhRvR...2b3204S}
}

@ARTICLE{Carr2026,
       author = {{Carr}, Bernard and {Iovino}, Antonio J. and {Perna}, Gabriele and {Vaskonen}, Ville and {Veerm{\"a}e}, Hardi},
        title = "{Primordial black holes: constraints, potential evidence and prospects}",
      journal = {arXiv e-prints},
         year = 2026,
        month = jan,
          eid = {arXiv:2601.06024},
        pages = {arXiv:2601.06024},
          doi = {10.48550/arXiv.2601.06024},
archivePrefix = {arXiv},
       eprint = {2601.06024},
 primaryClass = {astro-ph.CO},
       adsurl = {https://ui.adsabs.harvard.edu/abs/2026arXiv260106024C}
}

@ARTICLE{Kocevski2025,
       author = {{Kocevski}, Dale D. and {Finkelstein}, Steven L. and {Barro}, Guillermo and {et.al.}},
        title = "{The Rise of Faint, Red Active Galactic Nuclei at z > 4: A Sample of Little Red Dots in the JWST Extragalactic Legacy Fields}",
      journal = {\apj},
         year = 2025,
        month = jun,
       volume = {986},
       number = {2},
          eid = {126},
        pages = {126},
          doi = {10.3847/1538-4357/adbc7d},
archivePrefix = {arXiv},
       eprint = {2404.03576},
 primaryClass = {astro-ph.GA},
       adsurl = {https://ui.adsabs.harvard.edu/abs/2025ApJ...986..126K}
}

@ARTICLE{Lodato_etal,
       author = {{Lodato}, Giuseppe and {Natarajan}, Priyamvada},
        title = "{Supermassive black hole formation during the assembly of pre-galactic discs}",
      journal = {\mnras},
         year = 2006,
        month = oct,
       volume = {371},
       number = {4},
        pages = {1813-1823},
          doi = {10.1111/j.1365-2966.2006.10801.x},
archivePrefix = {arXiv},
       eprint = {astro-ph/0606159},
 primaryClass = {astro-ph},
       adsurl = {https://ui.adsabs.harvard.edu/abs/2006MNRAS.371.1813L}
}

@ARTICLE{Begelman_etal,
       author = {{Begelman}, Mitchell C. and {Volonteri}, Marta and {Rees}, Martin J.},
        title = "{Formation of supermassive black holes by direct collapse in pre-galactic haloes}",
      journal = {\mnras},
         year = 2006,
        month = jul,
       volume = {370},
       number = {1},
        pages = {289-298},
          doi = {10.1111/j.1365-2966.2006.10467.x},
archivePrefix = {arXiv},
       eprint = {astro-ph/0602363},
 primaryClass = {astro-ph},
       adsurl = {https://ui.adsabs.harvard.edu/abs/2006MNRAS.370..289B}
}

@ARTICLE{Bromm_etal,
       author = {{Bromm}, Volker and {Loeb}, Abraham},
        title = "{Formation of the First Supermassive Black Holes}",
      journal = {\apj},
         year = 2003,
        month = oct,
       volume = {596},
       number = {1},
        pages = {34-46},
          doi = {10.1086/377529},
archivePrefix = {arXiv},
       eprint = {astro-ph/0212400},
 primaryClass = {astro-ph},
       adsurl = {https://ui.adsabs.harvard.edu/abs/2003ApJ...596...34B}
}

@ARTICLE{Hennebelle,
       author = {{Hennebelle}, Patrick},
        title = "{Amplification and generation of turbulence during self-gravitating collapse}",
      journal = {\aap},
         year = 2021,
        month = nov,
       volume = {655},
          eid = {A3},
        pages = {A3},
          doi = {10.1051/0004-6361/202141650},
archivePrefix = {arXiv},
       eprint = {2109.01858},
 primaryClass = {astro-ph.GA},
       adsurl = {https://ui.adsabs.harvard.edu/abs/2021A&A...655A...3H}
}

@ARTICLE{Irshad+25,
       author = {{Irshad P}, Muhammed and {Bhat}, Pallavi and {Subramanian}, Kandaswamy and {Shukurov}, Anvar},
        title = "{Turbulent dynamos in a collapsing cloud}",
      journal = {arXiv e-prints},
         year = 2025,
        month = mar,
          eid = {arXiv:2503.19131},
        pages = {arXiv:2503.19131},
          doi = {10.48550/arXiv.2503.19131},
archivePrefix = {arXiv},
       eprint = {2503.19131},
 primaryClass = {astro-ph.GA},
       adsurl = {https://ui.adsabs.harvard.edu/abs/2025arXiv250319131I}
}

@ARTICLE{GR24,
       author = {{Graham}, Peter W. and {Ramani}, Harikrishnan},
        title = "{Constraints on dark matter from dynamical heating of stars in ultrafaint dwarfs. II. Substructure and the primordial power spectrum}",
      journal = {\prd},
         year = 2024,
        month = oct,
       volume = {110},
       number = {7},
          eid = {075012},
        pages = {075012},
          doi = {10.1103/PhysRevD.110.075012},
archivePrefix = {arXiv},
       eprint = {2404.01378},
 primaryClass = {hep-ph},
       adsurl = {https://ui.adsabs.harvard.edu/abs/2024PhRvD.110g5012G}
}

@ARTICLE{Choi_etal,
       author = {{Choi}, Jun-Hwan and {Shlosman}, Isaac and {Begelman}, Mitchell C.},
        title = "{Supermassive black hole formation at high redshifts via direct collapse in a cosmological context}",
      journal = {\mnras},
         year = 2015,
        month = jul,
       volume = {450},
       number = {4},
        pages = {4411-4423},
          doi = {10.1093/mnras/stv694},
       adsurl = {https://ui.adsabs.harvard.edu/abs/2015MNRAS.450.4411C}
}

@ARTICLE{Latif_etal,
       author = {{Latif}, M.~A. and {Schleicher}, D.~R.~G. and {Schmidt}, W. and {Niemeyer}, J.},
        title = "{Black hole formation in the early Universe}",
      journal = {\mnras},
         year = 2013,
        month = aug,
       volume = {433},
       number = {2},
        pages = {1607-1618},
          doi = {10.1093/mnras/stt834},
archivePrefix = {arXiv},
       eprint = {1304.0962},
 primaryClass = {astro-ph.CO},
       adsurl = {https://ui.adsabs.harvard.edu/abs/2013MNRAS.433.1607L}
}

@article{Benson+12,
    author = {Benson, Andrew J. and Farahi, Arya and Cole, Shaun and Moustakas, Leonidas A. and Jenkins, Adrian and Lovell, Mark and Kennedy, Rachel and Helly, John and Frenk, Carlos},
    title = {Dark matter halo merger histories beyond cold dark matter – I. Methods and application to warm dark matter},
    journal = {Monthly Notices of the Royal Astronomical Society},
    volume = {428},
    number = {2},
    pages = {1774-1789},
    year = {2012},
    month = {11},
    issn = {0035-8711},
    doi = {10.1093/mnras/sts159},
    url = {https://doi.org/10.1093/mnras/sts159},
    eprint = {https://academic.oup.com/mnras/article-pdf/428/2/1774/3285851/sts159.pdf},
}

@ARTICLE{FF25,
       author = {{Fakhry}, Saeed and {Firouzjaee}, Javad T.},
        title = "{Ultradense Dark Matter Halos with Poisson Noise from Stellar-mass Primordial Black Holes}",
      journal = {\apj},
         year = 2025,
        month = aug,
       volume = {989},
       number = {1},
          eid = {116},
        pages = {116},
          doi = {10.3847/1538-4357/adf21a},
archivePrefix = {arXiv},
       eprint = {2502.00914},
 primaryClass = {astro-ph.CO},
       adsurl = {https://ui.adsabs.harvard.edu/abs/2025ApJ...989..116F}
}

@ARTICLE{GI_Review,
       author = {{Kratter}, Kaitlin and {Lodato}, Giuseppe},
        title = "{Gravitational Instabilities in Circumstellar Disks}",
      journal = {\araa},
         year = 2016,
        month = sep,
       volume = {54},
        pages = {271-311},
          doi = {10.1146/annurev-astro-081915-023307},
archivePrefix = {arXiv},
       eprint = {1603.01280},
 primaryClass = {astro-ph.SR},
       adsurl = {https://ui.adsabs.harvard.edu/abs/2016ARA&A..54..271K}
}

@ARTICLE{BH_review,
       author = {{Balbus}, Steven A. and {Hawley}, John F.},
        title = "{Instability, turbulence, and enhanced transport in accretion disks}",
      journal = {Reviews of Modern Physics},
         year = 1998,
        month = jan,
       volume = {70},
       number = {1},
        pages = {1-53},
          doi = {10.1103/RevModPhys.70.1},
       adsurl = {https://ui.adsabs.harvard.edu/abs/1998RvMP...70....1B}
}

@ARTICLE{Flock+12,
       author = {{Flock}, M. and {Henning}, Th. and {Klahr}, H.},
        title = "{Turbulence in Weakly Ionized Protoplanetary Disks}",
      journal = {\apj},
         year = 2012,
        month = dec,
       volume = {761},
       number = {2},
          eid = {95},
        pages = {95},
          doi = {10.1088/0004-637X/761/2/95},
archivePrefix = {arXiv},
       eprint = {1210.4669},
 primaryClass = {astro-ph.EP},
       adsurl = {https://ui.adsabs.harvard.edu/abs/2012ApJ...761...95F}
}

@ARTICLE{PS,
       author = {{Press}, William H. and {Schechter}, Paul},
        title = "{Formation of Galaxies and Clusters of Galaxies by Self-Similar Gravitational Condensation}",
      journal = {\apj},
         year = 1974,
        month = feb,
       volume = {187},
        pages = {425-438},
          doi = {10.1086/152650},
       adsurl = {https://ui.adsabs.harvard.edu/abs/1974ApJ...187..425P}
}

@BOOK{Frank_etal_book,
       author = {{Frank}, J. and {King}, A.~R. and {Raine}, D.~J.},
        title = "{Accretion power in astrophysics}",
         year = 1985,
publisher = "{Cambridge University Press}",
       adsurl = {https://ui.adsabs.harvard.edu/abs/1985apa..book.....F}
}

@ARTICLE{Loeb1993,
       author = {{Loeb}, Abraham},
        title = "{Cosmological Formation of Quasar Black Holes}",
      journal = {\apj},
         year = 1993,
        month = feb,
       volume = {403},
        pages = {542},
          doi = {10.1086/172224},
       adsurl = {https://ui.adsabs.harvard.edu/abs/1993ApJ...403..542L}
}

@ARTICLE{Omukai+2005,
       author = {{Omukai}, K. and {Tsuribe}, T. and {Schneider}, R. and {Ferrara}, A.},
        title = "{Thermal and Fragmentation Properties of Star-forming Clouds in Low-Metallicity Environments}",
      journal = {\apj},
         year = 2005,
        month = jun,
       volume = {626},
       number = {2},
        pages = {627-643},
          doi = {10.1086/429955},
archivePrefix = {arXiv},
       eprint = {astro-ph/0503010},
 primaryClass = {astro-ph},
       adsurl = {https://ui.adsabs.harvard.edu/abs/2005ApJ...626..627O}
}

@article{Jangra_etal_2025,
doi = {10.1088/1475-7516/2025/08/006},
url = {https://doi.org/10.1088/1475-7516/2025/08/006},
year = {2025},
month = {aug},
publisher = {IOP Publishing},
volume = {2025},
number = {08},
pages = {006},
author = {Jangra, Pratibha and Gaggero, Daniele and Kavanagh, Bradley J. and Diego, J.M.},
title = {The cosmic history of Primordial Black Hole accretion and its uncertainties},
journal = {Journal of Cosmology and Astroparticle Physics},
}

@article{Ricotti_2008,
   title={Effect of Primordial Black Holes on the Cosmic Microwave Background and Cosmological Parameter Estimates},
   volume={680},
   ISSN={1538-4357},
   url={http://dx.doi.org/10.1086/587831},
   DOI={10.1086/587831},
   number={2},
   journal={The Astrophysical Journal},
   publisher={American Astronomical Society},
   author={Ricotti, Massimo and Ostriker, Jeremiah P. and Mack, Katherine J.},
   year={2008},
   month=jun, pages={829–845} }

@article{Ricotti_2007,
   title={Bondi Accretion in the Early Universe},
   volume={662},
   ISSN={1538-4357},
   url={http://dx.doi.org/10.1086/516562},
   DOI={10.1086/516562},
   number={1},
   journal={The Astrophysical Journal},
   publisher={American Astronomical Society},
   author={Ricotti, Massimo},
   year={2007},
   month=jun, pages={53–61} }

@BOOK{Draine_book,
       author = {{Draine}, Bruce T.},
        title = "{Physics of the Interstellar and Intergalactic Medium}",
         year = 2011,
publisher = "{Princeton University Press}",
       adsurl = {https://ui.adsabs.harvard.edu/abs/2011piim.book.....D}
}

@article{OHSUGI201844,
title = {Bondi–Hoyle–Lyttleton accretion flow revisited: Numerical simulation of unstable flow},
journal = {Astronomy and Computing},
volume = {25},
pages = {44-51},
year = {2018},
issn = {2213-1337},
doi = {https://doi.org/10.1016/j.ascom.2018.08.005},
url = {https://www.sciencedirect.com/science/article/pii/S2213133716301378},
author = {Y. Ohsugi},
}

@ARTICLE{Umemura_Fukue_1994PASJ...46..567U,
       author = {{Umemura}, Masayuki and {Fukue}, Jun},
        title = "{Cosmological Spherical Accretion via External Radiation Drag}",
      journal = {\pasj},
         year = 1994,
        month = dec,
       volume = {46},
       number = {6},
        pages = {567-574},
          doi = {10.1093/pasj/46.6.567},
       adsurl = {https://ui.adsabs.harvard.edu/abs/1994PASJ...46..567U}
}

@ARTICLE{Umemura_Loeb_Turner_1993,
       author = {{Umemura}, Masayuki and {Loeb}, Abraham and {Turner}, Edwin L.},
        title = "{Early Cosmic Formation of Massive Black Holes}",
      journal = {\apj},
         year = 1993,
        month = dec,
       volume = {419},
        pages = {459},
          doi = {10.1086/173499},
archivePrefix = {arXiv},
       eprint = {astro-ph/9303004},
 primaryClass = {astro-ph},
       adsurl = {https://ui.adsabs.harvard.edu/abs/1993ApJ...419..459U}
}

@INPROCEEDINGS{Bromm_Loeb_2003,
       author = {{Bromm}, Volker and {Loeb}, Abraham},
        title = "{The First Sources of Light}",
    booktitle = {The Emergence of Cosmic Structure},
         year = 2003,
       editor = {{Holt}, Stephen H. and {Reynolds}, Christopher S.},
       series = {American Institute of Physics Conference Series},
       volume = {666},
        month = may,
    publisher = {AIP},
        pages = {73-84},
          doi = {10.1063/1.1581773},
archivePrefix = {arXiv},
       eprint = {astro-ph/0301406},
 primaryClass = {astro-ph},
       adsurl = {https://ui.adsabs.harvard.edu/abs/2003AIPC..666...73B}
}

@inbook{Haiman_2012,
   title={The Formation of the First Massive Black Holes},
   ISBN={9783642323621},
   ISSN={0067-0057},
   url={http://dx.doi.org/10.1007/978-3-642-32362-1_6},
   DOI={10.1007/978-3-642-32362-1_6},
   booktitle={The First Galaxies},
   publisher={Springer Berlin Heidelberg},
   author={Haiman, Zoltán},
   year={2012},
   month=sep, pages={293–341} }

@ARTICLE{Ito_Omukai_2024,
       author = {{Ito}, Mana and {Omukai}, Kazuyuki},
        title = "{First star formation in extremely early epochs}",
      journal = {\pasj},
         year = 2024,
        month = aug,
       volume = {76},
       number = {4},
        pages = {850-862},
          doi = {10.1093/pasj/psae054},
archivePrefix = {arXiv},
       eprint = {2405.10073},
 primaryClass = {astro-ph.GA},
       adsurl = {https://ui.adsabs.harvard.edu/abs/2024PASJ...76..850I}
}

@article{Edgar_2004,
   title={A review of Bondi–Hoyle–Lyttleton accretion},
   volume={48},
   ISSN={1387-6473},
   url={http://dx.doi.org/10.1016/j.newar.2004.06.001},
   DOI={10.1016/j.newar.2004.06.001},
   number={10},
   journal={New Astronomy Reviews},
   publisher={Elsevier BV},
   author={Edgar, Richard},
   year={2004},
   month=sep, pages={843–859} }

@BOOK{Mo_White,
       author = {{Mo}, Houjun and {van den Bosch}, Frank C. and {White}, Simon},
        title = "{Galaxy Formation and Evolution}",
         year = 2010,
publisher = "{Cambridge University Press}",
          doi = {10.1017/CBO9780511807244},
       adsurl = {https://ui.adsabs.harvard.edu/abs/2010gfe..book.....M}
}

@ARTICLE{Hoyle_Lyt_1939,
       author = {{Hoyle}, F. and {Lyttleton}, R.~A.},
        title = "{The effect of interstellar matter on climatic variation}",
      journal = {Proceedings of the Cambridge Philosophical Society},
         year = 1939,
        month = jan,
       volume = {35},
       number = {3},
        pages = {405},
          doi = {10.1017/S0305004100021150},
       adsurl = {https://ui.adsabs.harvard.edu/abs/1939PCPS...35..405H}
}

@ARTICLE{Bondi_1952,
       author = {{Bondi}, H.},
        title = "{On spherically symmetrical accretion}",
      journal = {\mnras},
         year = 1952,
        month = jan,
       volume = {112},
        pages = {195},
          doi = {10.1093/mnras/112.2.195},
       adsurl = {https://ui.adsabs.harvard.edu/abs/1952MNRAS.112..195B}
}

@article{Tsel_2010,
   title={Relative velocity of dark matter and baryonic fluids and the formation of the first structures},
   volume={82},
   ISSN={1550-2368},
   url={http://dx.doi.org/10.1103/PhysRevD.82.083520},
   DOI={10.1103/physrevd.82.083520},
   number={8},
   journal={Physical Review D},
   publisher={American Physical Society (APS)},
   author={Tseliakhovich, Dmitriy and Hirata, Christopher},
   year={2010},
   month=oct }

@Inbook{Carr_Green,
author="Carr, Bernard J.
and Green, Anne M.",
editor="Byrnes, Christian
and Franciolini, Gabriele
and Harada, Tomohiro
and Pani, Paolo
and Sasaki, Misao",
title="The History of Primordial Black Holes",
bookTitle="Primordial Black Holes",
year="2025",
publisher="Springer Nature Singapore",
address="Singapore",
pages="3--33",
isbn="978-981-97-8887-3",
doi="10.1007/978-981-97-8887-3_1",
url="https://doi.org/10.1007/978-981-97-8887-3_1"
}

@ARTICLE{LVKGW231123,
author = {{The LIGO Scientific Collaboration} and {the Virgo Collaboration} and {the KAGRA Collaboration} and {Abac} et al, A.~G.},
title = "{GW231123: A Binary Black Hole Merger with Total Mass 190─265 M$_{{\ensuremath{\odot}}}$}",
      journal = {\apjl},
         year = 2025,
        month = nov,
       volume = {993},
       number = {1},
          eid = {L25},
        pages = {L25},
          doi = {10.3847/2041-8213/ae0c9c},
archivePrefix = {arXiv},
       eprint = {2507.08219},
 primaryClass = {astro-ph.HE},
       adsurl = {https://ui.adsabs.harvard.edu/abs/2025ApJ...993L..25A}
}

@ARTICLE{Bert,
       author = {{Bertschinger}, E.},
        title = "{Self-similar secondary infall and accretion in an Einstein-de Sitter universe}",
      journal = {\apjs},
         year = 1985,
        month = may,
       volume = {58},
        pages = {39-65},
          doi = {10.1086/191028},
       adsurl = {https://ui.adsabs.harvard.edu/abs/1985ApJS...58...39B}
}

@ARTICLE{Bogdan+,
       author = {{Bogd{\'a}n}, {\'A}kos and {Goulding}, Andy D. and {Natarajan}, Priyamvada and {Kov{\'a}cs}, Orsolya E. and {Tremblay}, Grant R. and {Chadayammuri}, Urmila and {Volonteri}, Marta and {Kraft}, Ralph P. and {Forman}, William R. and {Jones}, Christine and {Churazov}, Eugene and {Zhuravleva}, Irina},
        title = "{Evidence for heavy-seed origin of early supermassive black holes from a z {\ensuremath{\approx}} 10 X-ray quasar}",
      journal = {Nature Astronomy},
         year = 2024,
        month = jan,
       volume = {8},
       number = {1},
        pages = {126-133},
          doi = {10.1038/s41550-023-02111-9},
archivePrefix = {arXiv},
       eprint = {2305.15458},
 primaryClass = {astro-ph.GA},
       adsurl = {https://ui.adsabs.harvard.edu/abs/2024NatAs...8..126B}
}

@ARTICLE{Kovacs+,
       author = {{Kov{\'a}cs}, Orsolya E. and {Bogd{\'a}n}, {\'A}kos and {Natarajan}, Priyamvada and {Werner}, Norbert and {Azadi}, Mojegan and {Volonteri}, Marta and {Tremblay}, Grant R. and {Chadayammuri}, Urmila and {Forman}, William R. and {Jones}, Christine and {Kraft}, Ralph P.},
        title = "{A Candidate Supermassive Black Hole in a Gravitationally Lensed Galaxy at Z {\ensuremath{\approx}} 10}",
      journal = {\apjl},
         year = 2024,
        month = apr,
       volume = {965},
       number = {2},
          eid = {L21},
        pages = {L21},
          doi = {10.3847/2041-8213/ad391f},
archivePrefix = {arXiv},
       eprint = {2403.14745},
 primaryClass = {astro-ph.GA},
       adsurl = {https://ui.adsabs.harvard.edu/abs/2024ApJ...965L..21K}
}

@ARTICLE{Volonteri25,
       author = {{Volonteri}, Marta},
        title = "{Theoretical Modelling of Early Massive Black Holes}",
      journal = {arXiv e-prints},
         year = 2025,
        month = oct,
          eid = {arXiv:2510.04599},
        pages = {arXiv:2510.04599},
          doi = {10.48550/arXiv.2510.04599},
archivePrefix = {arXiv},
       eprint = {2510.04599},
 primaryClass = {astro-ph.GA},
       adsurl = {https://ui.adsabs.harvard.edu/abs/2025arXiv251004599V}
}

@ARTICLE{Dayal24,
       author = {{Dayal}, Pratika},
        title = "{Exploring a primordial solution for early black holes detected with JWST}",
      journal = {\aap},
         year = 2024,
        month = oct,
       volume = {690},
          eid = {A182},
        pages = {A182},
          doi = {10.1051/0004-6361/202451481},
archivePrefix = {arXiv},
       eprint = {2407.07162},
 primaryClass = {astro-ph.GA},
       adsurl = {https://ui.adsabs.harvard.edu/abs/2024A&A...690A.182D}
}

@ARTICLE{Harikane+,
       author = {{Harikane}, Yuichi and {Zhang}, Yechi and {Nakajima}, Kimihiko and {Ouchi}, Masami and {Isobe}, Yuki and {Ono}, Yoshiaki and {Hatano}, Shun and {Xu}, Yi and {Umeda}, Hiroya},
        title = "{A JWST/NIRSpec First Census of Broad-line AGNs at z = 4-7: Detection of 10 Faint AGNs with M $_{BH}$ {}10$^{6}$-{}10$^{8}$ M $_{{\ensuremath{\odot}}}$ and Their Host Galaxy Properties}",
      journal = {\apj},
         year = 2023,
        month = dec,
       volume = {959},
       number = {1},
          eid = {39},
        pages = {39},
          doi = {10.3847/1538-4357/ad029e},
archivePrefix = {arXiv},
       eprint = {2303.11946},
 primaryClass = {astro-ph.GA},
       adsurl = {https://ui.adsabs.harvard.edu/abs/2023ApJ...959...39H}
}

@ARTICLE{Ubler+,
       author = {{{\"U}bler}, Hannah and {Maiolino}, Roberto and {Curtis-Lake}, Emma and {P{\'e}rez-Gonz{\'a}lez}, Pablo G. and {Curti}, Mirko and {Perna}, Michele and {Arribas}, Santiago and {Charlot}, St{\'e}phane and {Marshall}, Madeline A. and {D'Eugenio}, Francesco and {Scholtz}, Jan and {Bunker}, Andrew and {Carniani}, Stefano and {Ferruit}, Pierre and {Jakobsen}, Peter and {Rix}, Hans-Walter and {Rodr{\'\i}guez Del Pino}, Bruno and {Willott}, Chris J. and {Boeker}, Torsten and {Cresci}, Giovanni and {Jones}, Gareth C. and {Kumari}, Nimisha and {Rawle}, Tim},
        title = "{GA-NIFS: A massive black hole in a low-metallicity AGN at z {\ensuremath{\sim}} 5.55 revealed by JWST/NIRSpec IFS}",
      journal = {\aap},
         year = 2023,
        month = sep,
       volume = {677},
          eid = {A145},
        pages = {A145},
          doi = {10.1051/0004-6361/202346137},
archivePrefix = {arXiv},
       eprint = {2302.06647},
 primaryClass = {astro-ph.GA},
       adsurl = {https://ui.adsabs.harvard.edu/abs/2023A&A...677A.145U}
}

@ARTICLE{Maiolino+,
       author = {{Maiolino}, Roberto and {Scholtz}, Jan and {Curtis-Lake}, Emma and {Carniani}, Stefano and {Baker}, William and {de Graaff}, Anna and {Tacchella}, Sandro and {{\"U}bler}, Hannah and {D'Eugenio}, Francesco and {Witstok}, Joris and {Curti}, Mirko and {Arribas}, Santiago and {Bunker}, Andrew J. and {Charlot}, St{\'e}phane and {Chevallard}, Jacopo and {Eisenstein}, Daniel J. and {Egami}, Eiichi and {Ji}, Zhiyuan and {Jones}, Gareth C. and {Lyu}, Jianwei and {Rawle}, Tim and {Robertson}, Brant and {Rujopakarn}, Wiphu and {Perna}, Michele and {Sun}, Fengwu and {Venturi}, Giacomo and {Williams}, Christina C. and {Willott}, Chris},
        title = "{JADES: The diverse population of infant black holes at 4 < z < 11: Merging, tiny, poor, but mighty}",
      journal = {\aap},
         year = 2024,
        month = nov,
       volume = {691},
          eid = {A145},
        pages = {A145},
          doi = {10.1051/0004-6361/202347640},
archivePrefix = {arXiv},
       eprint = {2308.01230},
 primaryClass = {astro-ph.GA},
       adsurl = {https://ui.adsabs.harvard.edu/abs/2024A&A...691A.145M}
}

@article{Delos_Silk,
    author = {Delos, M Sten and Silk, Joseph},
    title = {Ultradense dark matter haloes accompany primordial black holes},
    journal = {Monthly Notices of the Royal Astronomical Society},
    volume = {520},
    number = {3},
    pages = {4370-4375},
    year = {2023},
    month = {02},
    issn = {0035-8711},
    doi = {10.1093/mnras/stad356},
    url = {https://doi.org/10.1093/mnras/stad356},
    eprint = {https://academic.oup.com/mnras/article-pdf/520/3/4370/49288240/stad356.pdf},
}

@ARTICLE{Hu_Sugiyama,
       author = {{Hu}, Wayne and {Sugiyama}, Naoshi},
        title = "{Small-Scale Cosmological Perturbations: an Analytic Approach}",
      journal = {\apj},
         year = 1996,
        month = nov,
       volume = {471},
        pages = {542},
          doi = {10.1086/177989},
archivePrefix = {arXiv},
       eprint = {astro-ph/9510117},
 primaryClass = {astro-ph},
       adsurl = {https://ui.adsabs.harvard.edu/abs/1996ApJ...471..542H}
}

@ARTICLE{Lacey_Cole,
       author = {{Lacey}, Cedric and {Cole}, Shaun},
        title = "{Merger rates in hierarchical models of galaxy formation}",
      journal = {\mnras},
         year = 1993,
        month = jun,
       volume = {262},
       number = {3},
        pages = {627-649},
          doi = {10.1093/mnras/262.3.627},
       adsurl = {https://ui.adsabs.harvard.edu/abs/1993MNRAS.262..627L}
}

@ARTICLE{Planck18,
       author = {{Planck Collaboration} and {Aghanim}, N. and {Akrami}, Y. and {Ashdown et. al.}, M. },
        title = "{Planck 2018 results. VI. Cosmological parameters}",
      journal = {\aap},
         year = 2020,
        month = sep,
       volume = {641},
          eid = {A6},
        pages = {A6},
          doi = {10.1051/0004-6361/201833910},
archivePrefix = {arXiv},
       eprint = {1807.06209},
 primaryClass = {astro-ph.CO},
       adsurl = {https://ui.adsabs.harvard.edu/abs/2020A&A...641A...6P}
}

@article{Carr+,
title = {Cosmic conundra explained by thermal history and primordial black holes},
journal = {Physics of the Dark Universe},
volume = {31},
pages = {100755},
year = {2021},
issn = {2212-6864},
doi = {https://doi.org/10.1016/j.dark.2020.100755},
url = {https://www.sciencedirect.com/science/article/pii/S2212686420304684},
author = {Bernard Carr and Sébastien Clesse and Juan García-Bellido and Florian Kühnel},
}

@ARTICLE{COBE_FIRAS,
       author = {{Fixsen}, D.~J. and {Cheng}, E.~S. and {Gales}, J.~M. and {Mather}, J.~C. and {Shafer}, R.~A. and {Wright}, E.~L.},
        title = "{The Cosmic Microwave Background Spectrum from the Full COBE FIRAS Data Set}",
      journal = {\apj},
         year = 1996,
        month = dec,
       volume = {473},
        pages = {576},
          doi = {10.1086/178173},
archivePrefix = {arXiv},
       eprint = {astro-ph/9605054},
 primaryClass = {astro-ph},
       adsurl = {https://ui.adsabs.harvard.edu/abs/1996ApJ...473..576F}
}

@ARTICLE{Nakama_Carr_Silk,
       author = {{Nakama}, Tomohiro and {Carr}, Bernard and {Silk}, Joseph},
        title = "{Limits on primordial black holes from {\ensuremath{\mu}} distortions in cosmic microwave background}",
      journal = {\prd},
         year = 2018,
        month = feb,
       volume = {97},
       number = {4},
          eid = {043525},
        pages = {043525},
          doi = {10.1103/PhysRevD.97.043525},
archivePrefix = {arXiv},
       eprint = {1710.06945},
 primaryClass = {astro-ph.CO},
       adsurl = {https://ui.adsabs.harvard.edu/abs/2018PhRvD..97d3525N}
}

@ARTICLE{Bringmann+,
       author = {{Bringmann}, Torsten and {Croon}, Djuna and {Sevillano Mu{\~n}oz}, Sergio},
        title = "{Updated constraints on the primordial power spectrum at sub-Mpc scales}",
      journal = {arXiv e-prints},
         year = 2025,
        month = jun,
          eid = {arXiv:2506.20704},
        pages = {arXiv:2506.20704},
          doi = {10.48550/arXiv.2506.20704},
archivePrefix = {arXiv},
       eprint = {2506.20704},
 primaryClass = {astro-ph.CO},
       adsurl = {https://ui.adsabs.harvard.edu/abs/2025arXiv250620704B}
}

@ARTICLE{BL01,
       author = {{Barkana}, R. and {Loeb}, A.},
        title = "{In the beginning: the first sources of light and the reionization of the universe}",
      journal = {\physrep},
         year = 2001,
        month = jul,
       volume = {349},
       number = {2},
        pages = {125-238},
          doi = {10.1016/S0370-1573(01)00019-9},
archivePrefix = {arXiv},
       eprint = {astro-ph/0010468},
 primaryClass = {astro-ph},
       adsurl = {https://ui.adsabs.harvard.edu/abs/2001PhR...349..125B}
}

@ARTICLE{Wenzer+,
       author = {{Qin}, Wenzer and {Kumar}, Soubhik and {Natarajan}, Priyamvada and {Weiner}, Neal},
        title = "{Not-quite-primordial black holes}",
      journal = {arXiv e-prints},
         year = 2025,
        month = jun,
          eid = {arXiv:2506.13858},
        pages = {arXiv:2506.13858},
          doi = {10.48550/arXiv.2506.13858},
archivePrefix = {arXiv},
       eprint = {2506.13858},
 primaryClass = {astro-ph.CO},
       adsurl = {https://ui.adsabs.harvard.edu/abs/2025arXiv250613858Q}
}

@ARTICLE{Bird+,
       author = {{Bird}, Simeon and {Peiris}, Hiranya V. and {Viel}, Matteo and {Verde}, Licia},
        title = "{Minimally parametric power spectrum reconstruction from the Lyman {\ensuremath{\alpha}} forest}",
      journal = {\mnras},
         year = 2011,
        month = may,
       volume = {413},
       number = {3},
        pages = {1717-1728},
          doi = {10.1111/j.1365-2966.2011.18245.x},
archivePrefix = {arXiv},
       eprint = {1010.1519},
 primaryClass = {astro-ph.CO},
       adsurl = {https://ui.adsabs.harvard.edu/abs/2011MNRAS.413.1717B}
}

@ARTICLE{Fernandez+,
       author = {{Fernandez}, M.~A. and {Bird}, Simeon and {Ho}, Ming-Feng},
        title = "{Cosmological constraints from the eBOSS Lyman-{\ensuremath{\alpha}} forest using the PRIYA simulations}",
      journal = {\jcap},
         year = 2024,
        month = jul,
       volume = {2024},
       number = {7},
          eid = {029},
        pages = {029},
          doi = {10.1088/1475-7516/2024/07/029},
archivePrefix = {arXiv},
       eprint = {2309.03943},
 primaryClass = {astro-ph.CO},
       adsurl = {https://ui.adsabs.harvard.edu/abs/2024JCAP...07..029F}
}

\end{document}